\documentclass{aa}  

\usepackage{graphicx}
\usepackage{amsmath}	% Advanced maths commands
\usepackage{amssymb}	% Extra maths symbols
\usepackage{txfonts}
\usepackage{hyperref}
\usepackage{breakurl}
\usepackage{color}

\begin{document} 

   \title{On the incidence of planet candidates in open clusters and a planet confirmation\thanks{Based on observations collected with the 3.6~m Telescope (La Silla Observatory, ESO, Chile) using the HARPS instrument (programs ID: 091.C-0438, 092.C-0282, and 094.C-0297).}}

   %\subtitle{I. Overviewing the $\kappa$-mechanism}

   \titlerunning{Incidence of planet candidates in open clusters}

   \author{I.~C.~Le\~ao\inst{1}
          \and B.~L.~Canto~Martins\inst{1,2}
          \and S.~Alves\inst{3}
          \and G.~Pereira~de~Oliveira\inst{1}
          \and C.~Cort\'es\inst{4,5}
          \and A.~Brucalassi\inst{6}
          \and C.~H.~F.~Melo\inst{7}
          \and D.~B.~de~Freitas\inst{8}
          \and L.~Pasquini\inst{6}
          \and J.~R.~de~Medeiros\inst{1}
          }

         %1
   \institute{Universidade Federal do Rio Grande do Norte, Natal, Brazil\\
              \email{izan@dfte.ufrn.br}
         %2
         \and
         Observatoire de Gen\`eve, Universit\'e de Gen\`eve, 51 Ch. des Maillettes, 1290 Sauverny, Switzerland
         %3
         \and Universidade Federal do Rec\^oncavo da Bahia, Centro de Ci\^encias Exatas e Tecnol\'ogicas,  Av. Rui Barbosa, 710, Cruz das Almas, BA, 44380-000, Brazil
         %4
         \and Departamento de F\'isica, Facultad de Ciencias B\'asicas, Universidad Metropolitana de la Educaci\'on, Av. Jos\'e Pedro Alessandri 774, 7760197 Nu\~noa, Santiago, Chile
         %5
         \and Millennium Institute of Astrophysics (MAS), Santiago, Chile
         %6
         \and European Southern Observatory (ESO), Karl-Schwarzschild-Stra{\ss}e 2, 85748 Garching bei M\"unchen, Germany
         %7
         \and ESO, Casilla 19001, Santiago 19, Chile
         %8
         \and Departamento de F\'isica, Universidade Federal do Cear\'a, Caixa Postal 6030, Campus do Pici, 60455-900 Fortaleza, Cear\'a, Brazil
             }

   \date{Received ...; accepted ...}

% \abstract{}{}{}{}{} 
% 5 {} token are mandatory
 
  \abstract
   {Detecting exoplanets in clusters of different ages is a powerful tool for understanding a number of open questions, such as how the occurrence rate of planets depends on stellar metallicity, on mass, or on stellar environment.}
   {We present the first results of our HARPS long-term radial velocity (RV) survey which aims to discover exoplanets around intermediate-mass (between $\sim$ 2 and 6~M$_{\odot}$) evolved stars in open clusters.}
   {We selected 826 bona fide HARPS observations of 114 giants from an initial list of 29 open clusters and computed the half peak-to-peak variability of the HARPS RV measurements, namely $\Delta RV/2$, for each target, to search for the best planet-host candidates. We also performed time series analysis for a few targets with enough observations to search for orbital solutions.}
   {Although we attempted to rule out the presence of binaries on the basis of previous surveys, we detected 14 new binary candidates in our sample, most of them identified from a comparison between HARPS and CORAVEL data. We also suggest 11 new planet-host candidates based on a relation between the stellar surface gravity and $\Delta RV/2$. Ten of the candidates have less than 3~M$_{\odot}$, showing evidence of a low planet occurrence rate for massive stars. One of the planet-host candidates and one of the binary candidates show very clear RV periodic variations, allowing us to confirm the discovery of a new planet and to compute the orbital solution for the binary. The planet is IC~4651~9122b, with a minimum mass of $m\sin i = 6.3$~M$_{\rm J}$ and a semi-major axis $a = 2.0$~AU. The binary companion is NGC~5822~201B, with a very low minimum mass of $m\sin i = 0.11$~M$_{\odot}$ and a semi-major axis $a = 6.5$~AU, which is comparable to the Jupiter distance to the Sun.}
   {}


   \maketitle
%
%-------------------------------------------------------------------

%%%%%%%%%%%%%%%%%%%%%%%%%%%%%%%%%%%%%%%%%%%%%%%%%%%%%%%%%%%%%%%%%%%%%%%%%%%%
\section{Introduction} \label{introduction}
%%%%%%%%%%%%%%%%%%%%%%%%%%%%%%%%%%%%%%%%%%%%%%%%%%%%%%%%%%%%%%%%%%%%%%%%%%%

After the pioneering discovery of the giant planet orbiting 51 Peg by \citet{may95}, two decades ago, the literature\footnote{\url{http://exoplanet.eu/}} reports to date the discovery of more than  3700 planets, in about 2800 planetary systems.
Solar stars in the field host the vast majority of these exoplanets. The characteristics of field stars may represent a drawback for our capability to derive precise conclusions to very basic questions. For example, more than 70\% of the known planets orbit stars with masses $M_* < 1.30$~M$_{\odot}$. Our understanding of planet formation as a function of the mass of the host star and of the stellar environments is therefore still poorly understood.
In addition, it has been observed that main sequence stars hosting giant planets are metal-rich \citep{gon97,san04}, while evolved stars hosting giant planets are likely not (\citealt{pas07}, see, however, for different conclusions \citealt{jon16}). There is no clear explanation for this discrepancy, and several competing scenarios have been proposed, including stellar pollution acting on main-sequence stars \citep[e.g.,][]{lau97}, a planet formation (core-accretion) mechanism favoring the birth of planets around metal-rich stars
\citep{pol96}, and an effect of stellar migration (radial mixing) in the Galactic disk \citep{hay09}.

Open cluster stars formed simultaneously from a single molecular cloud with uniform physical properties, and thus have the same age, chemical composition and galactocentric distance. As a result, these are valuable testbeds for studying how the planet occurrence rate depends on stellar mass and environment.
Furthermore, comparing homogeneous sets of open-cluster stars with and without planets is an ideal method for determining whether the presence of a planetary companion alters the chemical composition of the host stars \citep[e.g.,][]{isr09}.

The number of planetary mass companions discovered around open cluster stars is rapidly growing, amounting to date to 25~planets. Two hot-Jupiters and a massive outer planet in the Praesepe open cluster \citep{qui12,mal16}, a hot-Jupiter in the Hyades \citep{qui14}, two sub-Neptune planets in NGC~6811 \citep{mei13}, five Jupiter-mass planets in~M67 \citep{bru14,bru16,bru17}, a Neptune-sized planet transiting an M4.5~dwarf in the Hyades \citep{man16,dav16},  three Earth-to-Neptune-sized planets around a mid-K dwarf in the Hyades \citep{man17b}, a Neptune-sized planet orbiting an M~dwarf in Praesepe \citep{obe16} and eight planets from~K2 campaigns \citep{pop16,bar16,lib16,man17a,cur18}, have been recently reported. Three planet candidates were also announced in the~M67 field \citep{nar16}, although all the host stars appear to be non-members. Previous radial velocity studies focusing on evolved stars revealed a giant planet around one of the Hyades clump giants \citep{sat07} and a substellar-mass object in NGC~2423 \citep{lov07}. These studies confirm that giant planets around open cluster stars exist and can probably migrate in a dense cluster environment. \citet{mei13} found that the properties and occurrence rate of low mass planets are the same in open clusters and field stars. Finally, the radial velocity (RV) measurements of~M67 show that the occurrence rate of giant planets is compatible with that observed in field stars ($\sim16\%$), albeit with an excess of hot jupiters in this cluster \citep{bru16}.

These studies demonstrate the wealth of information which can be gained from open clusters, but have so far been limited to solar mass stars.
It is necessary to extend the work to a broader range of stellar masses and ages for a better understanding about the planet occurrence rate related with the mass, environment, and chemical composition of the host stars.
However, more massive hot stars show very few and broad spectral lines, so cool stars in the red giant region are excellent candidates for extending these works to higher masses \citep{set04,joh07}.

Over the past three years we have carried out a search for massive planets around 152 evolved stars belonging to  29 open clusters. From these clusters we selected  114 targets with the best quality data and with a minimum of two observations per target, as described in Sect.~\ref{finalsample}.
These targets were relatively well studied for duplicity, and also with good constraints for mass, composition and age determinations. Our survey aims to estimate the planet occurrence rate of intermediate-mass late-type giant stars in young and intermediate-age open clusters. This paper provides an overview (as made in \citealt{pas12} for M67) of the stellar sample and the observations, discussing the clusters' characteristics and the RV distribution of the stars, and highlighting the most likely planetary host candidates.
The paper is structured as follows. The observations, sample selection, and methods used in our analysis are described in Sect.~\ref{methods}. Several results are presented in Sect.~\ref{results}, including a detailed overview of the data we have collected so far, combined with observations of other programs, and the discovery of a new planet. Finally, our conclusions are stated in Sect.~\ref{conclusions}.

%%%%%%%
\begin{table}
   \caption{Number of observed stars ($N_{\rm obj}$) of our original list of 29 open clusters and total number of HARPS observations$^a$ ($N_{\rm obs}$) carried out by our and other programs for each cluster.}
   {\centering
   \begin{tabular}{c c c | c c c}
     \hline\hline
Cluster & $N_{\rm obj}$ & $N_{\rm obs}$ & Cluster & $N_{\rm obj}$ & $N_{\rm obs}$ \\
     \hline
IC 2714 & 8 & 217 & NGC 2972 & 2 & 7 \\
IC 4651 & 13 & 150 & NGC 3114 & 7 & 89 \\
IC 4756 & 13 & 52 & NGC 3532 & 6 & 45 \\
Melotte 71 & 6 & 11 & NGC 3680 & 6 & 32 \\
NGC 1662 & 2 & 4 & NGC 3960 & 3 & 6 \\
NGC 2204 & 8 & 20 & NGC 4349 & 3 & 98 \\
NGC 2251 & 3 & 3 & NGC 5822 & 11 & 96 \\
NGC 2324 & 3 & 3 & NGC 6067 & 3 & 13 \\
NGC 2345 & 4 & 8 & NGC 6134 & 9 & 14 \\
NGC 2354 & 8 & 22 & NGC 6208 & 2 & 6 \\
NGC 2355 & 1 & 1 & NGC 6281 & 2 & 7 \\
NGC 2477 & 10 & 24 & NGC 6425 & 2 & 6 \\
NGC 2506 & 6 & 8 & NGC 6494 & 2 & 8 \\
NGC 2818 & 3 & 8 & NGC 6633 & 4 & 22 \\
NGC 2925 & 2 & 14 &  &  \\
     \hline\hline
     \end{tabular} \\ }
     \small
     \vspace{0.1in}
     {\bf Note.}\\
     $^a$Total of 994 observations of 152 stars from which we selected our final sample of 826 effective observations of  114 stars, as described in Sect.~\ref{finalsample}.\vspace{0.1cm}\\
  \label{tabclus}
\end{table}
%%%%%%%

%%%%%%%%%%%%%%%%%%%%%%%%%%%%%%%%%%%%%%%%%%%%%%%%%%%%%%%%%%%%%%%%%%%%%%%%%%%%
\section{Working sample, observations, and methods}\label{methods}
%%%%%%%%%%%%%%%%%%%%%%%%%%%%%%%%%%%%%%%%%%%%%%%%%%%%%%%%%%%%%%%%%%%%%%%%%%%%

The stellar sample was selected from \citet{mer08}, who determined cluster membership and binaries using CORAVEL.
The stellar B and V magnitudes, $B_{\rm mag}$ and $V_{\rm mag}$, were obtained from the Simbad\footnote{\url{http://simbad.u-strasbg.fr/simbad/}} database, from which we also computed the $(B-V)$ color index.
Absolute magnitudes, $M_V$, were estimated from $V_{\rm mag}$ and the cluster distance modulus obtained from \citet{kha05} and from the WEBDA\footnote{\url{http://webda.physics.muni.cz/}} cluster database \citep{mer95}.
Both $(B-V)$ and $M_V$ were corrected for reddening, $E(B-V)$, from \citet{wu09}.
Cluster ages were taken from \citet{wu09} and metallicities from \citet{wu09} and \citet{hei14}.

The main cluster selection criteria were the age of the cluster (between 0.02 and $\sim$2~Gyr, with turnoff masses $\gtrsim 2$~M$_{\odot}$) and the apparent magnitude of the giant stars (brighter than $V_{\rm mag} \sim 14$~mag).
We then rejected cool, bright stars with $(B-V) > 1.4$ as these are known to be RV unstable \citep[e.g.,][]{hek06}.
Known binaries and non-members were removed from the sample. It is important to note that we pick up only the clusters with at least 2 giant members each and that the chosen open clusters span a rather narrow metallicity range (about $-0.2 < {\rm [Fe/H]} < 0.2$). Seven clusters were in common with \citet{lov07}. To first order we assumed that all giants in a given cluster have the same mass, which is approximately the mass at the main-sequence turnoff.

%%%%%%%%%%%%%%%%%%%%%%%%%%%%%%%%%%%%%%%%%%%%%%%
\subsection{HARPS observations}\label{harpsobs}

The {\em High Accuracy Radial velocity Planet Searcher\footnote{\url{http://www.eso.org/sci/facilities/lasilla/instruments/harps.html}}} (HARPS; \citealt{may03}) is the planet hunter at the ESO 3.6 m telescope in La Silla. In high accuracy mode (HAM) it has an aperture on the sky of one arcsecond, and a resolving power of 115000. The spectral range covered is 380--680 nm. In addition to be exceptionally stable, HARPS achieves the highest precision using the simultaneous calibration principle: the spectrum of a calibration (Th-Ar) source is recorded simultaneously with the stellar spectrum, with a second optical fiber.
As a rule of thumb we can consider the precision of HARPS scales as $\epsilon RV \propto (S/N)^{-1}$, where $\epsilon RV$ is the RV photon-noise error and $S/N$ is the signal-to-noise ratio. An $\epsilon RV$ of a few m/s ($\sim$10~m~s$^{-1}$ or better) is possible with limited $S/N$ observations, namely at least $S/N \sim 10$.
In practice, our observed HARPS spectra have typically a peak $S/N$ of 10--20 for the faintest stars and of 50--100 for the brightest ones.
HARPS is equipped with a very powerful pipeline \citep{may03} that provides on-line RV measurements, which are computed by cross correlating the stellar spectrum with a numerical template mask. This on-line pipeline also provides an associated $\epsilon RV$. For all of our stars, irrespective of the spectral type and luminosity, we used the solar template (G2V) mask. 

Between April 4th, 2013 and April 1st, 2015 we obtained  500 observations of 152 targets with HARPS spread over our  29 open clusters.
We then combined these data with  494 more HARPS observations of stars that were in our sample and collected using the same G2V mask, all available in the ESO Archive.
This provided a total of  994 observations for these targets obtained with a decade-long baseline, from 2005 to 2015,
from which we selected our final sample of  826 effective observations of  114 stars, as described in Sect.~\ref{finalsample}.
Table~\ref{tabclus} lists the stars by cluster, with the initial number of objects and observations considered in this work.

%%%%%%%
\begin{figure}
  \centering 
   \includegraphics[width=\columnwidth]{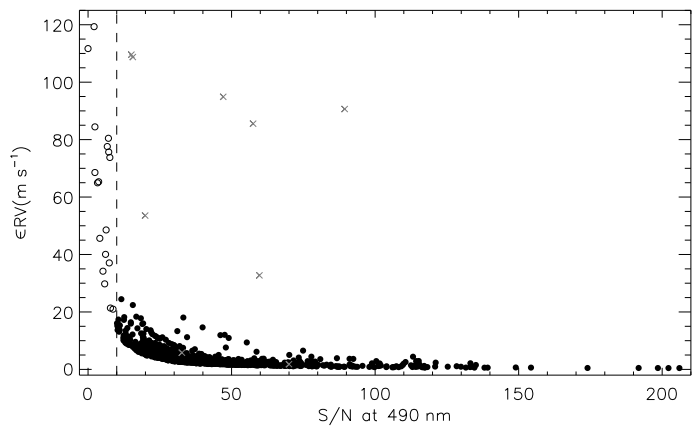}
   \caption{Photon noise error in the measurement of the CCF center of single observations vs. $S/N$ at 490 nm for our sample of open cluster targets observed with HARPS. The overall RV errors are typically distributed around 1--4~m/s and range from 42~cm/s to 119~m/s. The overall $S/N$ distribution peaks around 20--40 and ranges from 2.4 to 223. Black filled circles are the selected data, and gray crosses are the discarded ones, all identified from a visual inspection of the CCFs. Black open circles depict the data with $S/N < 10$ that were also discarded and this threshold is represented by the vertical dashed line.}
   \label{snrfig}
\end{figure}
%%%%%%%

%%%%%%%
\begin{figure}
  \centering 
   \includegraphics[width=\columnwidth]{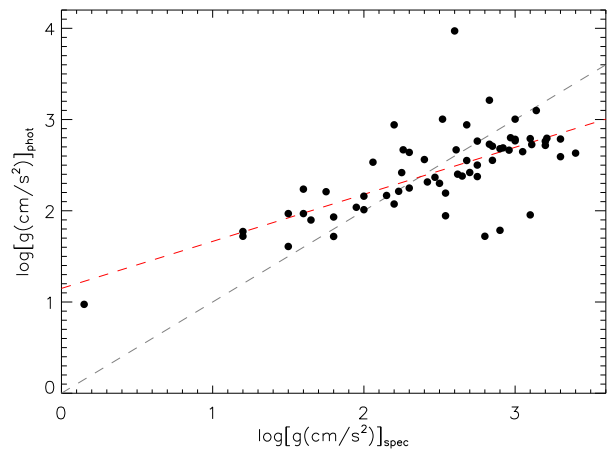}
   \caption{Comparison between spectroscopic and photometric $\log g$. The 1:1 relashionship is shown by the black dashed line. A systematic trend is illustrated by the red dashed line. The final photometric $\log g$ values were corrected for this trend.}
   \label{loggphot}
\end{figure}
%%%%%%%

%%%%%%%%%%%%%%%%%%%%%%%%%%%%%%%%%%%%%%%%%%%%%%%%%%%%%%%%%%%%%%%%%%%%%%%%%%%%
\begin{table*}
   \caption{Analysis of $\Delta RV/2$ for the final sample described in Sect.~\ref{finalsample}.}
   {\centering
   \begin{tabular}{l c c c | c c c c c c}
   \hline\hline
Object & $V_{\rm mag}$ & $\log g$       & $RV_{\rm M08}$ & $N_{\rm eff}$$^a$ & $t_{\rm span}$ & $\left<{RV}\right>$ & $\Delta RV/2$ & $\Delta RV_{\rm H-C}$$^b$ & flag$^c$ \\
       & (mag)         & [cm s$^{-2}$]  & (km s$^{-1}$)  &                   & (d)            & (km s$^{-1}$)       &  (m s$^{-1}$) &  (m s$^{-1}$)              &          \\
     \hline
IC 2714 5 & 11.10 & 2.70 & $-$14.53 & 25 & 3661 & $-$14.388 & 27.76 & 83 & \\
IC 2714 53 & 11.522 & 2.75 & $-$13.37 & 21 & 3198 & $-$13.269 & 45.34 & 42 & p(A) \\
IC 2714 87 & 11.395 & 2.62 & $-$13.23 & 23 & 3661 & $-$13.205 & 44.84 & $-$34 & \\
IC 2714 110 & 11.73 & 2.85 & $-$13.80 & 27 & 3661 & $-$13.771 & 42.00 & $-$31 & p(B) \\
IC 2714 121 & 10.80 & 2.10$^d$ & $-$13.37 & 21 & 1570 & $-$13.358 & 27.40 & $-$47 & \\
IC 2714 126 & 11.04 & 2.68$^d$ & $-$14.42 & 30 & 3661 & $-$14.130 & 24.95 & 230 & \\
IC 2714 190 & 11.32 & 2.55$^d$ & $-$13.60 & 27 & 3662 & $-$13.541 & 25.26 & 0 & \\
IC 2714 220 & 11.13 & 2.62$^d$ & $-$13.03 & 28 & 3560 & $-$13.374 & 69.12 & $-$403 & p(C) \\
IC 4651 6333 & 10.44 & 2.05$^d$ & $-$30.87 & 3 & 216 & $-$30.489 & 12.75 & 92 & \\
IC 4651 7646 & 10.363 & 2.61 & $-$31.18 & 7 & 3285 & $-$31.103 & 4.15 & $-$211 & \\
IC 4651 8540 & 10.894 & 2.26 & $-$30.36 & 25 & 3284 & $-$30.177 & 21.88 & $-$105 & \\
IC 4651 9025 & 10.90 & 2.90 & $-$30.46 & 26 & 3284 & $-$30.261 & 25.03 & $-$89 & \\
IC 4651 9122 & 10.7 & 2.52 & $-$30.58 & 51 & 3284 & $-$30.253 & 116.39 & 38 & p(D) \\
IC 4651 9791 & 10.44 & 2.23 & $-$31.44 & 7 & 3050 & $-$31.152 & 30.02 & 0 & \\
IC 4651 11218 & 11.09 & 3.00 & $-$30.40 & 2 & 20 & $-$31.113 & 1.54 & $-$1001 & B \\
IC 4651 12935 & 11.00 & 4.38$^d$ & $-$30.26 & 3 & 215 & $-$29.727 & 8.30 & 244 & \\
IC 4756 12 & 9.54 & 2.75 & $-$25.25 & 3 & 81 & $-$25.128 & 2.99 & $-$73 & \\
IC 4756 14 & 8.86 & 2.47 & $-$24.78 & 3 & 79 & $-$22.843 & 21.62 & 1743 & B \\
IC 4756 28 & 9.01 & 2.42 & $-$25.26 & 4 & 105 & $-$24.982 & 10.98 & 84 & \\
IC 4756 38 & 9.83 & 3.00 & $-$25.78 & 8 & 3087 & $-$25.650 & 5.12 & $-$65 & \\
IC 4756 42 & 9.46 & 3.21 & $-$24.92 & 2 & 16 & $-$24.719 & 1.37 & 7 & \\
IC 4756 44 & 9.77 & 3.30 & $-$26.01 & 6 & 2593 & $-$25.814 & 21.17 & 0 & p \\
IC 4756 49 & 9.46 & 2.83 & $-$25.40 & 4 & 110 & $-$25.164 & 10.06 & 41 & \\
IC 4756 52 & 8.06 & 3.10 & $-$25.21 & 4 & 136 & $-$25.132 & 48.81 & $-$117 & p \\
IC 4756 81 & 9.46 & 3.00 & $-$23.25 & 3 & 76 & $-$23.060 & 36.99 & $-$4 & p \\
IC 4756 101 & 9.36 & 3.20 & $-$25.74 & 3 & 81 & $-$25.592 & 4.42 & $-$47 & \\
IC 4756 109 & 9.05 & 3.30 & $-$25.25 & 4 & 111 & $-$24.693 & 14.80 & 362 & \\
IC 4756 125 & 9.36 & 3.11 & $-$24.85 & 3 & 81 & $-$24.751 & 5.36 & $-$95 & \\
IC 4756 164 & 9.27 & 3.40 & $-$25.51 & 4 & 106 & $-$25.294 & 7.93 & 23 & \\
Melotte 71 3 & 14.113 & 4.31$^d$ & +50.45 & 2 & 21 & +50.753 & 5.52 & 0 & \\
Melotte 71 19 & 11.880 & 2.62$^d$ & +49.64 & 2 & 21 & +50.244 & 8.27 & 301 & \\
Melotte 71 23 & 10.990 & 1.53$^d$ & +49.73 & 2 & 21 & +49.565 & 13.30 & $-$467 & \\
Melotte 71 121 & 12.800 & 2.69$^d$ & +50.91 & 2 & 22 & +51.185 & 6.60 & $-$28 & \\
Melotte 71 130 & 12.687 & 2.53$^d$ & +49.92 & 2 & 23 & +50.416 & 4.79 & 193 & \\
NGC 2204 1320 & 12.607 & 2.55$^d$ & +91.83 & 2 & 21 & +91.522 & 24.25 & $-$337 & \\
NGC 2204 2136 & 13.122 & 2.64$^d$ & +89.09 & 2 & 19 & +93.318 & 197.59 & 4199 & B \\
NGC 2204 2212 & 12.76 & 2.40$^d$ & +92.11 & 2 & 21 & +92.252 & 35.68 & 113 & \\
NGC 2204 3324 & 12.830 & 2.13$^d$ & +90.73 & 2 & 19 & +90.759 & 2.55 & 0 & \\
NGC 2204 3325 & 11.563 & $-$1.04$^d$ & +92.67 & 2 & 22 & +91.786 & 83.38 & $-$913 & B \\
NGC 2204 4137 & 11.97 & 2.82$^d$ & +91.13 & 2 & 23 & +92.709 & 2.99 & 1550 & B \\
NGC 2345 14 & 10.73 & 0.05$^d$ & +59.80 & 2 & 22 & +58.860 & 2.25 & $-$411 & \\
NGC 2345 43 & 10.70 & 0.37$^d$ & +58.82 & 2 & 22 & +58.492 & 7.87 & 201 & \\
NGC 2345 50 & 12.82 & $-$0.01$^d$ & +60.41 & 2 & 21 & +59.152 & 56.74 & $-$730 & B \\
NGC 2345 60 & 10.48 & 0.33$^d$ & +58.41 & 2 & 21 & +57.881 & 11.36 & 0 & \\
NGC 2354 66 & 11.73 & 1.74$^d$ & +34.08 & 3 & 35 & +34.281 & 15.48 & $-$115 & \\
NGC 2354 91 & 11.656 & 1.66$^d$ & +34.11 & 2 & 21 & +34.245 & 22.56 & $-$180 & \\
NGC 2354 125 & 11.73 & 1.73$^d$ & +32.44 & 3 & 35 & +33.347 & 4.98 & 591 & \\
NGC 2354 152 & 12.870 & 2.20$^d$ & +34.25 & 2 & 20 & +34.566 & 0.17 & 0 & \\
NGC 2354 183 & 11.555 & 2.90 & +34.25 & 2 & 21 & +34.524 & 6.66 & $-$41 & \\
NGC 2354 205 & 11.13 & 2.80 & +33.73 & 2 & 23 & +34.148 & 26.10 & 102 & \\
NGC 2354 219 & 11.001 & 1.69$^d$ & +31.50 & 3 & 37 & +32.330 & 11.37 & 514 & \\
     \hline\hline
     \end{tabular} \\ }
     \small
     \vspace{0.1in}
     {\bf Notes.}\\
     $^a$ $N_{\rm eff}$ refers to the effective number of observations after averaging those collected within less than three days of time interval (see Sect.~\ref{finalsample}).\vspace{0.1cm}\\
     $^b$ $\Delta RV_{\rm H-C} = \left<{RV}\right> - RV_{\rm M08} - {\it Offset}$, where {\it Offset} is provided in Table~\ref{tabif}.\\
     $^c$ Flags are ``p'' for planet-host candidate, where ``(A)'', ``(B)'', etc we analyze in more detail (see Sect.~\ref{sectcandidates}); ``B'' for long-period binary; and ``[B]'' for short-period binary (see Sect.~\ref{binaries}).\vspace{0.1cm}\\
     $^d$ Photometric estimation (see Sect.~\ref{sectloggphot}).\\
  \label{tablemain}
\end{table*}

\setcounter{table}{1}
\begin{table*}
   \caption{Continued.}
   {\centering
   \begin{tabular}{l c c c | c c c c c c}
   \hline\hline
Object & $V_{\rm mag}$ & $\log g$       & $RV_{\rm M08}$ & $N_{\rm eff}$$^a$ & $t_{\rm span}$ & $\left<{RV}\right>$ & $\Delta RV/2$ & $\Delta RV_{\rm H-C}$$^b$ & flag$^c$ \\
       & (mag)         & [cm s$^{-2}$]  & (km s$^{-1}$)  &                   & (d)            & (km s$^{-1}$)       &  (m s$^{-1}$) &  (m s$^{-1}$)              &          \\
     \hline
NGC 2477 4004 & 10.811 & 2.30$^d$ & +7.05 & 4 & 87 & +7.609 & 24.54 & 402 & \\
NGC 2477 6254 & 10.853 & 2.63$^d$ & +8.86 & 4 & 84 & +9.021 & 47.57 & 0 & \\
NGC 2477 6288 & 11.39 & 2.57$^d$ & +8.86 & 3 & 85 & +8.916 & 4.42 & $-$104 & \\
NGC 2506 2212 & 11.9 & 1.75 & +83.56 & 2 & 21 & +83.548 & 11.28 & 0 & \\
NGC 2506 3254 & 11.12 & 1.30$^d$ & +83.17 & 2 & 21 & +82.452 & 92.53 & $-$706 & B \\
NGC 2818 3035 & 13.346 & 3.42$^d$ & +22.02 & 2 & 252 & +20.886 & 61.88 & 0 & p \\
NGC 2925 95 & 9.894 & 2.79$^d$ & +10.48 & 5 & 596 & +10.860 & 35.31 & 368 & \\
NGC 2925 108 & 9.94 & 3.26$^d$ & +9.39 & 6 & 697 & +9.400 & 21.17 & 0 & p \\
NGC 2972 3 & 12.12 & 2.18$^d$ & +20.14 & 2 & 363 & +19.812 & 50.44 & $-$487 & \\
NGC 2972 11 & 12.09 & 2.14$^d$ & +19.64 & 3 & 363 & +19.799 & 12.92 & 0 & \\
NGC 3114 6 & 7.69 & 1.2 & $-$1.43 & 14 & 3661 & $-$1.450 & 131.57 & $-$161 & \\
NGC 3114 150 & 8.00 & 1.8 & $-$2.19 & 8 & 697 & $-$1.253 & 58.73 & 796 & B \\
NGC 3114 170 & 7.32 & 1.5 & $-$1.95 & 4 & 600 & $-$2.225 & 31.46 & $-$417 & \\
NGC 3114 181 & 8.31 & 1.65 & $-$2.18 & 9 & 705 & $-$2.067 & 52.50 & $-$28 & \\
NGC 3114 238 & 8.49 & 1.6 & $-$1.72 & 8 & 706 & $-$1.571 & 18.83 & 7 & \\
NGC 3114 262 & 8.56 & 2.2 & $-$1.20 & 15 & 3667 & $-$1.059 & 63.81 & 0 & \\
NGC 3114 283 & 7.68 & 1.2 & $-$1.73 & 8 & 697 & $-$1.393 & 92.11 & 195 & \\
NGC 3532 19 & 7.702 & 2.65 & +2.94 & 5 & 722 & +3.851 & 26.85 & 660 & \\
NGC 3532 100 & 7.457 & 2.15 & +4.49 & 4 & 723 & +4.740 & 6.58 & 0 & \\
NGC 3532 122 & 8.189 & 2.60 & +3.34 & 5 & 722 & +3.479 & 34.88 & $-$110 & \\
NGC 3532 221 & 6.03 & 1.50 & +3.58 & 7 & 730 & +3.830 & 41.76 & 0 & \\
NGC 3532 596 & 7.869 & 2.25 & +2.50 & 5 & 723 & +5.531 & 104.11 & 2781 & B \\
NGC 3532 670 & 6.978 & 1.80 & +3.97 & 5 & 723 & +4.208 & 155.26 & $-$1 & \\
NGC 3680 13 & 10.78 & 2.68 & +1.48 & 5 & 620 & +1.472 & 37.09 & $-$140 & \\
NGC 3680 26 & 10.8 & 2.68 & +0.67 & 4 & 618 & +0.267 & 112.82 & $-$535 & p \\
NGC 3680 34 & 10.69 & 2.2 & +1.93 & 4 & 616 & +3.762 & 86.58 & 1700 & B \\
NGC 3680 41 & 10.886 & 2.40 & +1.28 & 6 & 721 & +1.601 & 59.98 & 188 & \\
NGC 3680 44 & 10.02 & 2.00 & +1.49 & 6 & 722 & +1.754 & 32.08 & 132 & \\
NGC 3680 53 & 10.7 & 2.30 & +1.11 & 6 & 719 & +1.240 & 59.58 & 0 & \\
NGC 3960 28 & 13.01 & 2.06 & $-$22.48 & 2 & 270 & $-$22.073 & 3.74 & 0 & \\
NGC 3960 44 & 14.86 & 2.46$^d$ & $-$21.42 & 3 & 269 & $-$20.315 & 13.54 & 698 & \\
NGC 4349 5 & 11.511 & 2.54 & $-$12.27 & 32 & 3663 & $-$11.971 & 26.92 & 19 & \\
NGC 4349 9 & 11.594 & 1.78$^d$ & $-$11.75 & 24 & 3025 & $-$11.669 & 47.53 & $-$199 & \\
NGC 4349 53 & 11.33 & 1.96$^d$ & $-$10.44 & 33 & 3663 & $-$10.160 & 46.89 & 0 & \\
NGC 5822 1 & 9.08 & 2.00 & $-$30.97 & 6 & 338 & $-$30.350 & 16.94 & 439 & \\
NGC 5822 6 & 10.78 & 2.95$^d$ & $-$29.50 & 4 & 113 & $-$29.346 & 16.81 & $-$27 & \\
NGC 5822 8 & 10.37 & 2.71$^d$ & $-$30.51 & 19 & 3284 & $-$29.479 & 150.47 & 849 & B \\
NGC 5822 102 & 10.84 & 3.20 & $-$29.77 & 5 & 141 & $-$29.588 & 44.25 & 0 & p \\
NGC 5822 201 & 10.26 & 2.85 & $-$27.90 & 16 & 2903 & $-$28.017 & 957.77 & $-$299 & [B] \\
NGC 5822 224 & 10.84 & 3.14 & $-$29.64 & 4 & 116 & $-$30.871 & 12.32 & $-$1413 & B \\
NGC 5822 240 & ~ & 1.95 & $-$29.46 & 5 & 282 & $-$29.209 & 18.88 & 70 & \\
NGC 5822 316 & 10.47 & 3.05 & $-$28.31 & 5 & 340 & $-$28.229 & 7.12 & $-$101 & \\
NGC 5822 348 & 10.97 & 2.96$^d$ & $-$29.06 & 4 & 118 & $-$29.177 & 3.34 & $-$298 & \\
NGC 5822 375 & 9.69 & 2.17$^d$ & $-$29.50 & 6 & 337 & $-$29.224 & 40.46 & 93 & \\
NGC 5822 443 & 9.72 & 2.18$^d$ & $-$29.25 & 6 & 336 & $-$28.972 & 30.88 & 96 & \\
NGC 6067 261 & 8.79 & 0.15 & $-$39.39 & 5 & 135 & $-$39.120 & 135.32 & 0 & \\
NGC 6067 298 & 8.47 & 1.35 & $-$39.74 & 2 & 55 & $-$45.603 & 48.87 & $-$6121 & B \\
NGC 6067 316 & 8.86 & 1.51$^d$ & $-$40.97 & 5 & 138 & $-$40.269 & 77.07 & 442 & \\
NGC 6134 62 & 11.892 & 2.72$^d$ & $-$26.02 & 3 & 120 & $-$26.018 & 5.32 & $-$48 & \\
NGC 6134 75 & 12.394 & 3.10 & $-$25.69 & 2 & 116 & $-$25.640 & 22.56 & 0 & \\
NGC 6134 129 & 12.53 & 2.83 & $-$25.95 & 3 & 117 & $-$25.360 & 18.12 & 540 & \\
NGC 6208 19 & 10.88 & 2.40$^d$ & $-$32.17 & 4 & 129 & $-$32.022 & 39.90 & 0 & \\
NGC 6208 31 & 11.60 & 2.98$^d$ & $-$32.83 & 2 & 115 & $-$32.549 & 11.67 & 133 & \\
NGC 6281 3 & 7.94 & 2.30 & $-$5.95 & 4 & 126 & $-$5.579 & 4.03 & 11 & \\
NGC 6281 4 & 8.16 & 2.50 & $-$5.21 & 3 & 115 & $-$4.850 & 16.12 & 0 & \\
NGC 6425 46 & 10.788 & 2.49$^d$ & $-$3.75 & 3 & 130 & $-$3.511 & 15.60 & 213 & \\
NGC 6425 61 & 10.75 & 2.54$^d$ & $-$3.19 & 3 & 130 & $-$3.164 & 10.08 & 0 & \\
NGC 6494 46 & 9.42 & 2.07$^d$ & $-$8.25 & 4 & 132 & $-$8.386 & 9.14 & $-$246 & \\
     \hline\hline
     \end{tabular} \\ }
\end{table*}

\setcounter{table}{1}
\begin{table*}
   \caption{Continued.}
   {\centering
   \begin{tabular}{l c c c | c c c c c c}
   \hline\hline
Object & $V_{\rm mag}$ & $\log g$       & $RV_{\rm M08}$ & $N_{\rm eff}$$^a$ & $t_{\rm span}$ & $\left<{RV}\right>$ & $\Delta RV/2$ & $\Delta RV_{\rm H-C}$$^b$ & flag$^c$ \\
       & (mag)         & [cm s$^{-2}$]  & (km s$^{-1}$)  &                   & (d)            & (km s$^{-1}$)       &  (m s$^{-1}$) &  (m s$^{-1}$)              &          \\
     \hline
NGC 6494 48 & 9.54 & 2.54 & $-$8.36 & 4 & 132 & $-$8.251 & 7.30 & 0 & \\
NGC 6633 100 & 8.31 & 2.75 & $-$28.98 & 4 & 127 & $-$28.740 & 12.34 & 13 & \\
NGC 6633 106 & 8.69 & 2.96 & $-$28.46 & 10 & 3084 & $-$28.372 & 13.04 & $-$140 & \\
NGC 6633 119 & 8.98 & 2.97 & $-$28.96 & 3 & 130 & $-$28.793 & 23.85 & $-$61 & \\
NGC 6633 126 & 8.77 & 2.92 & $-$29.27 & 3 & 136 & $-$29.042 & 9.43 & 0 & \\
     \hline\hline
     \end{tabular} \\ }
\end{table*}
%%%%%%%%%%%%%%%%%%%%%%%%%%%%%%%%%%%%%%%%%%%%%%%%%%%%%%%%%%%%%%%%%%%%%%%%%%%%

%%%%%%%
\begin{figure}
  \centering 
   \includegraphics[width=\columnwidth]{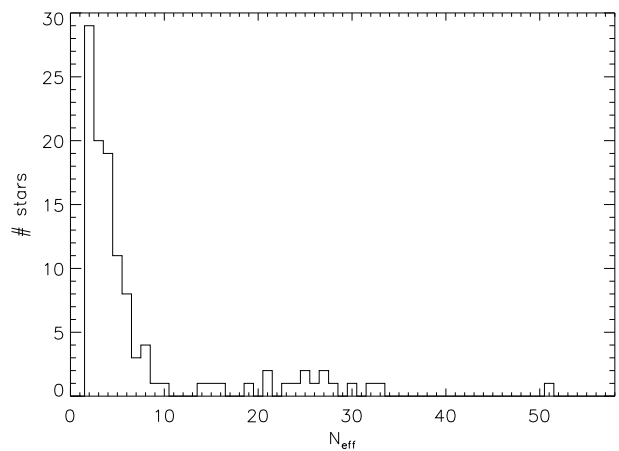}
   \caption{Distribution of the number of effective observations per target for our final sample of  114 targets with  826 effective observations described in Sect.~\ref{finalsample}.}
   \label{histnobs}
\end{figure}
%%%%%%%

%%%%%%%%%%%%%%%%%%%%%%%%%%%%%
\subsection{The final sample}
\label{finalsample}

The HARPS pipeline usually produces very good quality data from a fully automatic process. However, a visual inspection of the reduced data is
required to remove outliers caused by a variety of different issues (bad observation conditions, erroneous reduction, etc.).
Therefore, we performed a visual inspection of all the  994 HARPS observations described in Sect.~\ref{harpsobs} and, for a few cases, we re-reduced the data manually by using the offline tools of the HARPS pipeline to correct reduction issues.

We discarded  18 spectra with S/N ratio below 10 and  nine spectra with problematic cross-correlation function (CCF) shape.
Finally, we merged observations separated by less than 3~days by averaging all the reduced parameters (time, RV, bisector velocity span, and $S/N$).
This averaging tends to clean up short-period variations (likely produced by intrinsic stellar signal) and keeps periods longer than $\sim$10~d, which correspond, for the stellar mass range of our sample ($\sim$2--6~M$_{\odot}$), to planetary semi-major axes $\gtrsim 0.1$~AU (i.e., a threshold comparable to the typical radii of giant stars, below which inner planet orbits are not expected in our sample).
From the remaining data, we required at least two observations per target to analyze the RV distributions as described below. This final sample comprises a total of  826 effective observations of  114 targets.
Figure~\ref{snrfig} shows the distribution of $\epsilon RV$ vs. $S/N$ for our sample,
with objects belonging to different criteria in the refinement to final sample highlighted by different symbols.

%%%%%%%%%%%%%%%%%%%%%%%%%%%%%%%%%%%%%%%%
\subsection{Activity proxy measurements}
\label{actproxies}

The bisector velocity span (or simply bisector span) is a measurement of the CCF asymmetry
computed from the CCF bisector (which is the set of midpoints between the two sides of a CCF profile)
at the top and bottom of the CCF profile \citep[e.g.,][]{que01}.
It is a standard output of the HARPS pipeline and an important stellar activity proxy which is used to verify whether an RV variation is caused by intrinsic stellar variability (e.g., induced by chromospheric activity) rather than by orbital motion.
It has been demonstrated that, in the case of activity-induced RV variations, these correlate with the bisector span variations \citep[e.g.,][]{san02}.

Another important stellar activity proxy is the S~index, obtained from the emission in the core of the CaII H \& K spectral lines.
It is a dimensionless quantity typically measured from the total flux counts of two triangular passbands 1.09\AA ~wide centered at 3933.66\AA ~(the CaII K line) and at 3968.47\AA ~(the CaII H line) and normalized by the total flux of two continuum passbands 20\AA ~wide centered at 3901.07\AA ~(a pseudo blue filter) and 4001.07\AA ~(a pseudo red filter) \citep[e.g.,][]{sch09}. The S~index is, thus, defined as:
\begin{equation}
  \hspace{0.5cm} S_{\rm index} = \alpha\frac{H+K}{R+V},
\end{equation}
where $H$, $K$, $R$, and $V$ are the total flux counts of the pseudo-filters described above and $\alpha$ is a factor for instrumental calibration.
We measured this index by using the reduced HARPS spectra, which is also provided by the pipeline,
but no instrumental calibration was performed (ie., we assumed $\alpha = 1$) because we are only interested in the index variation.
This index, as the bisector span, should not correlate with RV if the RV variation has an orbital origin,
and it is reliable only for observations with high $S/N$ (e.g., $S/N \gtrsim 21$ at 400~nm; see Sect.~\ref{sectheplanet}).

%%%%%%%%%%%%%%%%%%%%%%%%%%%%%%%%%%%%%%%%%%%%%%%%%%%%
\subsection{Method to select planet-host candidates}
\label{selmethod}

A reasonable selection of planet-host candidates can be obtained by using the relation described in \citet{hek08}.
Based on a sample of K~giants, these authors found a trend where the RV semi-amplitude increases with the decreasing logarithm of stellar surface gravity, $\log g$.
This trend may arise from intrinsic RV variability induced by stellar oscillation \citep{kje95}, and was also observed by other groups \citep[e.g.,][]{set04}.
Planet-host candidates can be identified as those with RV lying noticeably above the trend.

\citet{tro16} followed this approach to search for exoplanets using data from APOGEE\footnote{\url{http://www.sdss.org/surveys/apogee/}} \citep{maj15}, with a pre-selection criterion quantified by their Eqs.~(25)--(27).
At first order,
these equations state that a planet-host candidate has RV semi-amplitudes above $3\times$ the trend level and above $3\times$ the typical RV error.
We therefore consider this criterion to pre-select our planet-host candidates.

To perform this analysis, we assumed that the half peak-to-peak difference between the available RV measurements, $\Delta RV/2$, represents the RV semi-amplitude. Of course, $\Delta RV/2$ may be biased for objects with a small number of observations.
We used  $\log g$ spectroscopic measurements provided in the PASTEL catalog \citep{sou10,sou16} when available,
and computed photometric values for the remaining targets by following the procedure described below.
New $\log g$ values computed from HARPS spectra will be provided in a forthcoming work (Canto Martins et al. 2018, in prep.).

%%%%%%%
\begin{figure}
  \centering 
   \includegraphics[width=\columnwidth]{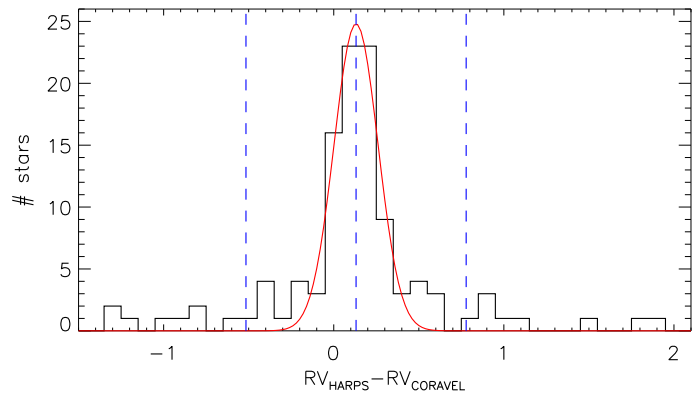}
   \caption{Distribution of the difference between the RV average obtained with HARPS and the RV average obtained with CORAVEL for each target.
                 The range is truncated for a better display; the whole distribution ranges from $-5.86$ to 4.16~km/s.
                 A gaussian fit is shown with its center and 5$\sigma$ range.}
   \label{histrvdiff}
\end{figure}
%%%%%%%

%%%%%%%
\begin{table}
   \caption{Summary of the selection of single star candidates for each cluster.}
   {\centering
   \begin{tabular}{c c c c c c c}
     \hline\hline
Cluster & $N_i$ & $N_f$ & {\it Offset} & $\sigma$   \\
            &            &             & (m s$^{-1}$) & (m s$^{-1}$) \\
     \hline
IC 2714 & 8 & 8 & 59 & 179 \\
IC 4651 & 8 & 7 & 289 & 149 \\
IC 4756 & 13 & 12 & 194 & 126 \\
Melotte 71 & 5 & 5 & 303 & 295 \\
NGC 1662 & -- & -- & -- & -- \\
NGC 2204 & 6 & 3 & 29 & 234 \\
NGC 2251 & -- & -- & -- & -- \\
NGC 2324 & -- & -- & -- & -- \\
NGC 2345 & 4 & 3 & $-$529 & 312 \\
NGC 2354 & 7 & 7 & 316 & 306 \\
NGC 2355 & -- & -- & -- & -- \\
NGC 2477 & 3 & 3 & 161 & 267 \\
NGC 2506 & 2 & 1 & $-$12 & -- \\
NGC 2818 & 1 & 1 & 353 & -- \\
NGC 2925 & 2 & 2 & 10 & 260 \\
NGC 2972 & 2 & 2 & 159 & 344 \\
NGC 3114 & 7 & 6 & 141 & 206 \\
NGC 3532 & 6 & 5 & 249 & 311 \\
NGC 3680 & 6 & 5 & 132 & 288 \\
NGC 3960 & 2 & 2 & 407 & 493 \\
NGC 4349 & 3 & 3 & 280 & 121 \\
NGC 5822 & 11 & 9 & 182 & 225 \\
NGC 6067 & 3 & 2 & 258 & 312 \\
NGC 6134 & 3 & 3 & 50 & 327 \\
NGC 6208 & 2 & 2 & 148 & 94 \\
NGC 6281 & 2 & 2 & 360 & 8 \\
NGC 6425 & 2 & 2 & 26 & 151 \\
NGC 6494 & 2 & 2 & 109 & 174 \\
NGC 6633 & 4 & 4 & 228 & 70 \\
     \hline\hline
     \end{tabular} \\ }
     \small
     \vspace{0.1in}
     {\bf Notes.}\\
     $N_i$ and $N_f$ refer to the number of objects before and after the removal of binaries.
     \textit{Offset} is the typical difference between HARPS and CORAVEL RVs, whereas $\sigma$ is the dispersion of this difference, both given for each cluster after the removal of binaries (see text for more information).
     \vspace{0.1cm}\\
  \label{tabif}
\end{table}
%%%%%%%

%%%%%%%%%%%%%%%%%%%%%%%%%%%%%%%%%%%%%%%%%%%%%%
\subsection{Photometric $\log(g)$ estimations}
\label{sectloggphot}

We estimated photometric $\log g$ values from a grid of isochrones of solar metallicity by using the CMD\footnote{\url{http://stev.oapd.inaf.it/cgi-bin/cmd}} Web Interface \citep[e.g.,][]{bre12,tan14,che14,che15}.
Each isochrone was traced on a grid $\log g\left((B-V),M_V\right)$ by using linear interpolation and the empty spaces between the isochrones were fulfilled by evolving a Laplace interpolation.
This single grid was used for the sake of simplicity, considering that the clusters have a metallicity around the solar value.
The central location of each target in the grid was used to get the theoretical $\log g$ value, which was set as an initial photometric $\log g$ estimation for the target.

After estimating an initial $\log g$ for all the targets, we plotted these values against the corresponding measured spectroscopic values and verified a systematic linear trend between them (see Fig.~\ref{loggphot}). We then corrected this trend to obtain the final photometric $\log g$ estimations.
The standard deviation between the photometric and spectroscopic $\log g$ values is $\sim$0.5~dex.
Thus, these photometric estimations can be used with caution for the purpose of our work, that is specifically the analysis of $\log g$ versus $\Delta RV/2$ described in Sect.~\ref{sectcandidates}.

%%%%%%%%%%%%%%%%%%%%%%%%%%%%%%%%%%%%%%%%%%%%%%%%%%%%%%%%%%%%
\subsection{Number of observations and time series analysis}
\label{secnobs}

We defined an arbitrary threshold of at least nine effective observations to perform time-series analysis of our planet-host and binary candidates. This is roughly a minimal requirement to obtain an orbital solution without ambiguity.
Figure~\ref{histnobs} shows the distribution of effective observations for the stars of our sample.
There are  94 stars with less than nine observations and  20 stars with at least nine observations.
The time-series analyses are presented in Sects.~\ref{sectimeseries} and \ref{sectheplanet} for the best planet-host candidates and for a binary candidate. An overall discussion of all the planet-host candidates is presented in Sect.~\ref{secmp}.

%%%%%%%
\begin{figure}
  \centering 
   \includegraphics[width=\columnwidth]{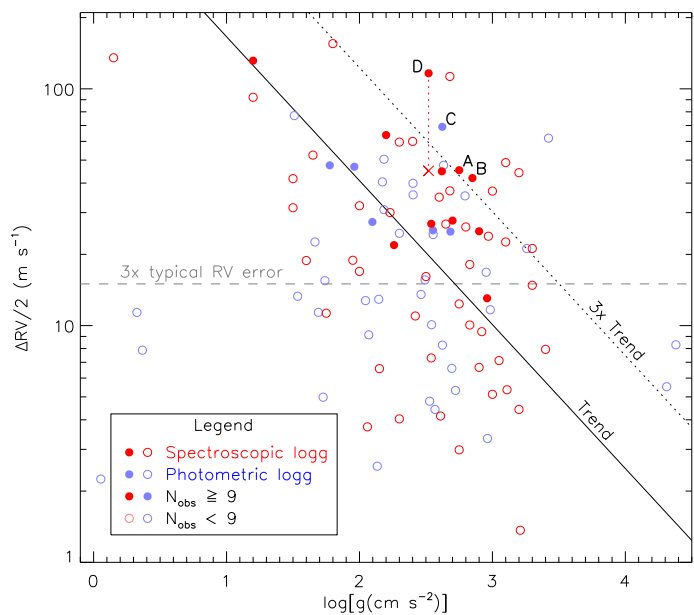}
   \caption{RV half peak-to-peak difference, $\Delta RV/2$, as a function of $\log g$, for our subsample of  101 single stars. Open circles stand for targets with a number of observations between two and eight, whereas filled circles represent the targets with at least nine observations. Red circles refer to the targets with spectroscopic $\log g$ measurements, whereas blue circles illustrate those with photometric measurements. The gray horizontal dashed line represents the 3$\times$ RV typical error level. The black solid line illustrates the linear fit of the data and the black dotted line is its $3\times$ level. The planet-host candidates lie in the upper right region encompassed within the dashed and the dotted lines. The candidates with at least nine observations are labeled A to D. The red cross illustrates the residual for target~D (IC~4651~9122) after removing the planet signal.}
   \label{loggvsrms}
\end{figure}
%%%%%%%

%%%%%%%
\begin{table*}
   \caption{Overview of planet and binary candidates each cluster.}
   {\centering
   %{\hspace{-0.3cm}
   \begin{tabular}{l c c c c c c c c}
     \hline\hline
Cluster & $\log(t)$ & $[Fe/H]$ & $M_{\rm TO}$  & $e(B-V)$ & $\mu$ & $N_{\rm *}$ & $N_{\rm p}$ & $N_{\rm b}$ \\
        & [yr]      & (dex)    & (M$_{\odot}$) & (mag)    & (mag)               &             &             \\
     \hline
NGC 3960 & 9.100 & $-$0.04 & 2.0 & 0.302 & 12.70 & 2 & 0 & 0 \\
NGC 2506 & 9.045 & $-$0.23 & 2.0 & 0.08 & 12.94 & 2 & 0 & 1 \\
NGC 3680 & 9.077 & $-$0.01 & 2.0 & 0.07 & 10.07 & 6 & 1 & 1 \\
NGC 6208 & 9.069 & $-$0.03 & 2.0 & 0.21 & 10.51 & 2 & 0 & 0 \\
IC 4651 & 9.057 & 0.12 & 2.1 & 0.12 & 10.11 & 8 & 1 & 1 \\
NGC 2204 & 8.896 & $-$0.32 & 2.2 & 0.085 & 12.36 & 6 & 0 & 3 \\
NGC 6134 & 8.968 & 0.11 & 2.2 & 0.38 & 10.98 & 3 & 0 & 0 \\
NGC 2355 & 8.850 & $-$0.05 & 2.4 & 0.12 & 12.08 & -- & -- & -- \\
NGC 5822 & 8.821 & 0.08 & 2.5 & 0.15 & 10.28 & 11 & 1 & 3 \\
NGC 2477 & 8.780 & 0.07 & 2.6 & 0.28 & 11.30 & 3 & 0 & 0 \\
IC 4756 & 8.699 & 0.02 & 2.7 & 0.19 & 9.01 & 13 & 3 & 1 \\
NGC 2324 & 8.650 & $-$0.22 & 2.8 & 0.127 & 13.30 & -- & -- & -- \\
NGC 2818 & 8.626 & $-$0.17 & 2.8 & 0.121 & 11.72 & 1 & 1 & 0 \\
NGC 6633 & 8.629 & $-$0.08 & 2.9 & 0.18 & 8.48 & 4 & 0 & 0 \\
IC 2714 & 8.542 & 0.02 & 3.1 & 0.34 & 11.52 & 8 & 3 & 0 \\
NGC 3532 & 8.492 & 0.00 & 3.2 & 0.04 & 8.61 & 6 & 0 & 1 \\
NGC 6281 & 8.497 & 0.06 & 3.3 & 0.15 & 8.93 & 2 & 0 & 0 \\
NGC 6494 & 8.477 & $-$0.04 & 3.3 & 0.36 & 10.11 & 2 & 0 & 0 \\
NGC 2251 & 8.427 & $-$0.09 & 3.4 & 0.19 & 11.21 & -- & -- & -- \\
Melotte 71 & 8.371 & $-$0.27 & 3.5 & 0.11 & 12.84 & 5 & 0 & 0 \\
NGC 1662 & 8.625 & 0.05 & 3.5 & 0.30 & 9.13 & -- & -- & -- \\
NGC 6425 & 7.346 & 0.09 & 3.7 & 0.40 & 10.69 & 2 & 0 & 0 \\
NGC 4349 & 8.315 & $-$0.07 & 3.8 & 0.38 & 12.87 & 3 & 0 & 0 \\
NGC 2354 & 8.126 &  & 4.4 & 0.29 & 13.79 & 7 & 0 & 0 \\
NGC 3114 & 8.093 & 0.05 & 4.7 & 0.08 & 10.05 & 7 & 0 & 1 \\
NGC 6067 & 8.076 & 0.14 & 4.8 & 0.40 & 12.00 & 3 & 0 & 1 \\
NGC 2972 & 7.968 & $-$0.07 & 5.2 & 0.343 & 12.63 & 2 & 0 & 0 \\
NGC 2345 & 7.853 &  & 5.9 & 0.68 & 13.87 & 4 & 0 & 1 \\
NGC 2925 & 7.850 &  & 5.9 & 0.08 & 9.69 & 2 & 1 & 0 \\
     \hline\hline
     \end{tabular} \\ }
     \small
     \vspace{0.1in}
     {\bf Note.}\\
     $N_*$ = number of stars in our sample; $N_p$ = number of planet-host candidates; $N_b$ = number of binaries.\\
  \label{tabcounts}
\end{table*}
%%%%%%%

%%%%%%%%%%%%%%%%%%%%%%%%%%%%%%%%%%%%%%%%%%%%%%%%%%%%%%%%%%%%%%%%%%%%%%%%%%%%
\section{Results}\label{results}
%%%%%%%%%%%%%%%%%%%%%%%%%%%%%%%%%%%%%%%%%%%%%%%%%%%%%%%%%%%%%%%%%%%%%%%%%%%%

Our final sample of  114 targets, obtained as described in Sect.~\ref{finalsample}, is listed in Table~\ref{tablemain}. The table includes the apparent visual magnitude ($V_{\rm mag}$), stellar surface gravity ($\log g$), and CORAVEL RV measurements ($RV_{\rm M08}$) from \citet{mer08}. Our data are the effective number of observations ($N_{\rm eff}$), time span ($t_{\rm span}$), the RV average ($\left<{RV}\right>$), the half peak-to-peak RV ($\Delta RV/2$), and the difference between the HARPS and CORAVEL RV values with respect to their offsets ($\Delta RV_{\rm H-C}$), these computed for each target. There is also a flag indicating the binary and planet-host candidates. The flag definitions and more details about the table parameters are discussed below.

%%%%%%%%%%%%%%%%%%%%%%%%%%%%%%%%%%%%%%%%%%%%%%
\subsection{Binary candidates}\label{binaries}

%%%%%%%
\begin{figure}
  \centering 
   \includegraphics[width=\columnwidth]{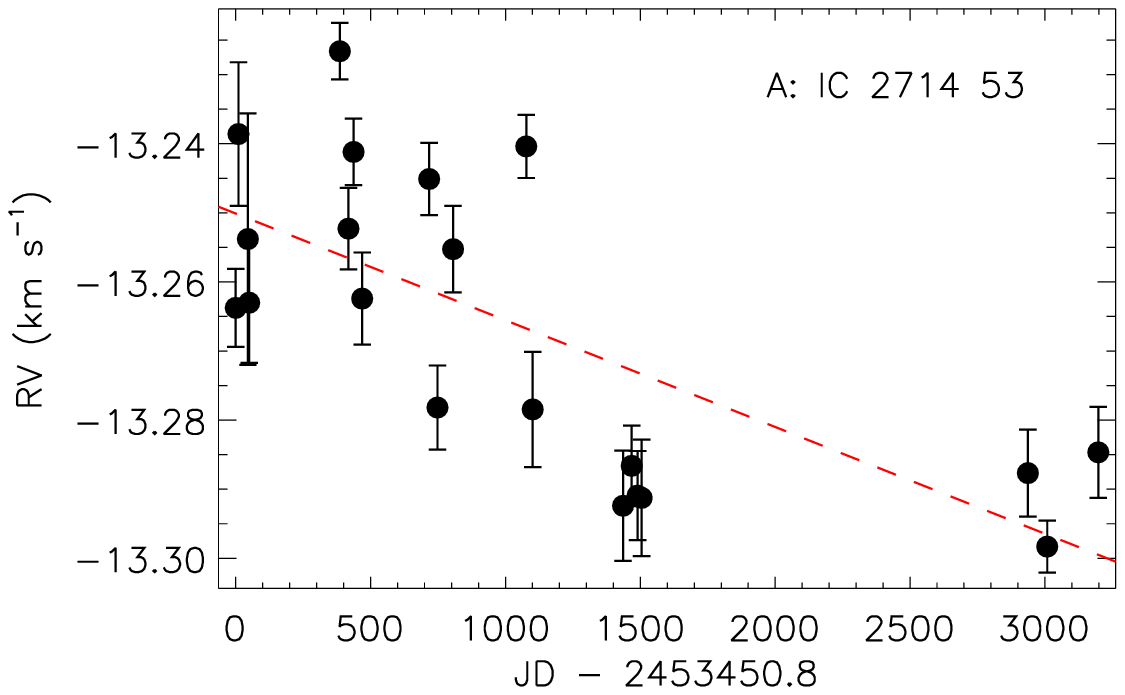}
   \includegraphics[width=\columnwidth]{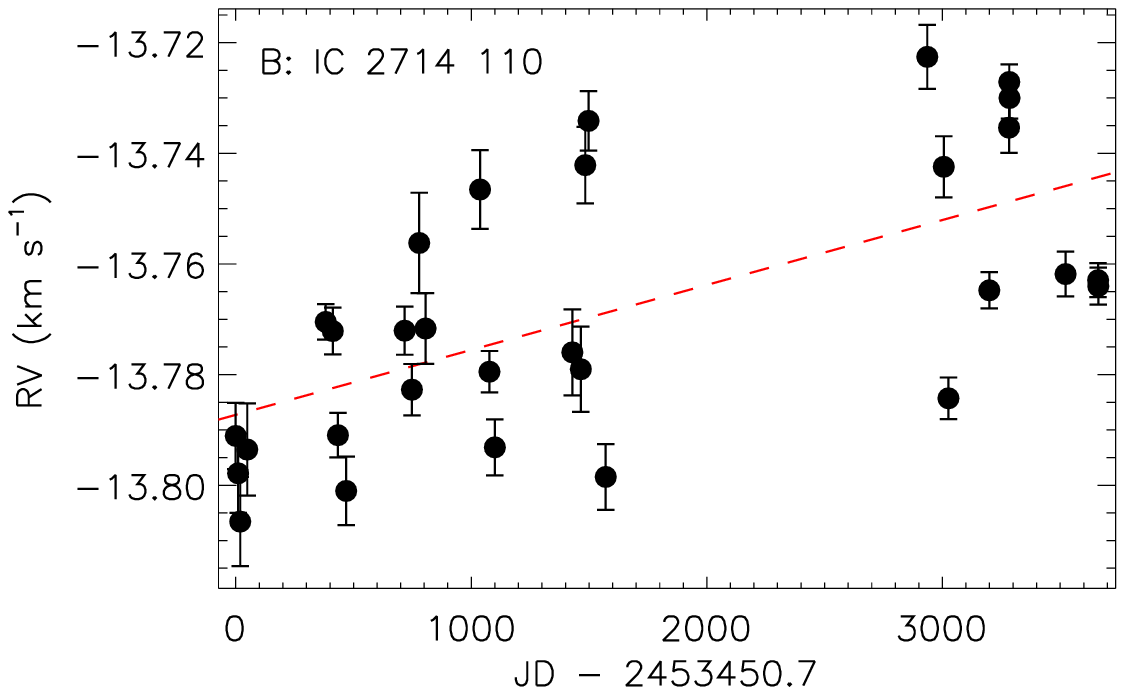}
   \includegraphics[width=\columnwidth]{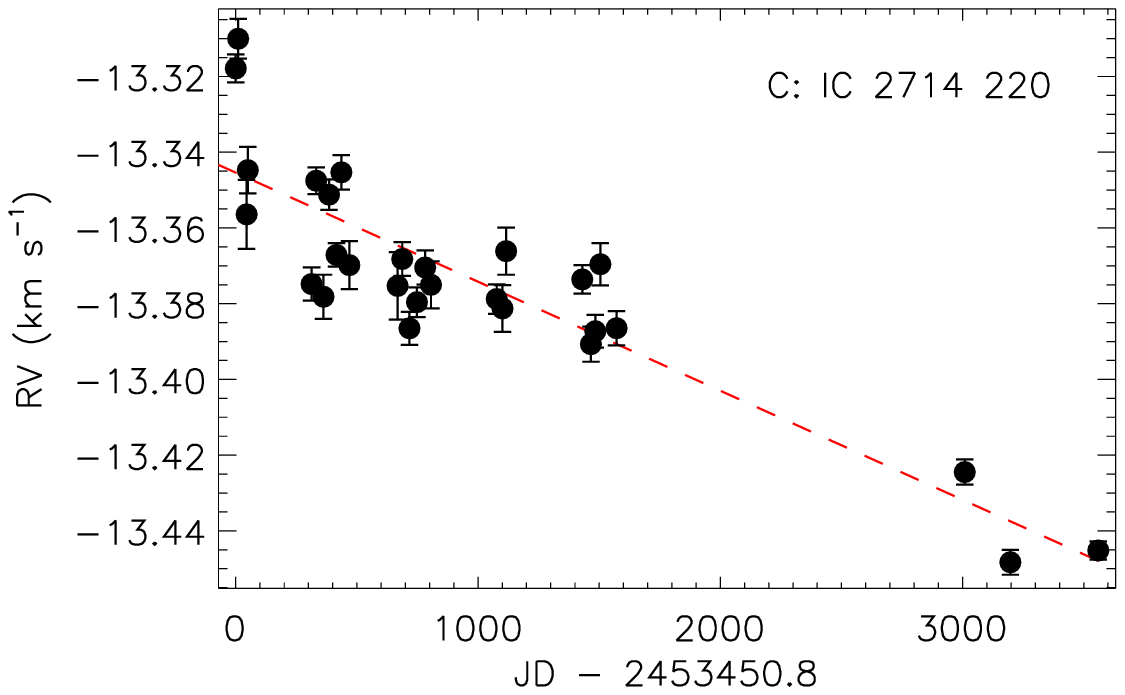}
   \caption{RV time series of the targets labeled A, B, and C in Fig.~\ref{loggvsrms}. All stars belong to the cluster IC~2714 and their RV data exhibit long-term RV variations, illustrated by the red dashed lines.}
   \label{timeseries}
\end{figure}
%%%%%%%

%%%%%%%
\begin{figure}
  \centering 
   \includegraphics[width=\columnwidth]{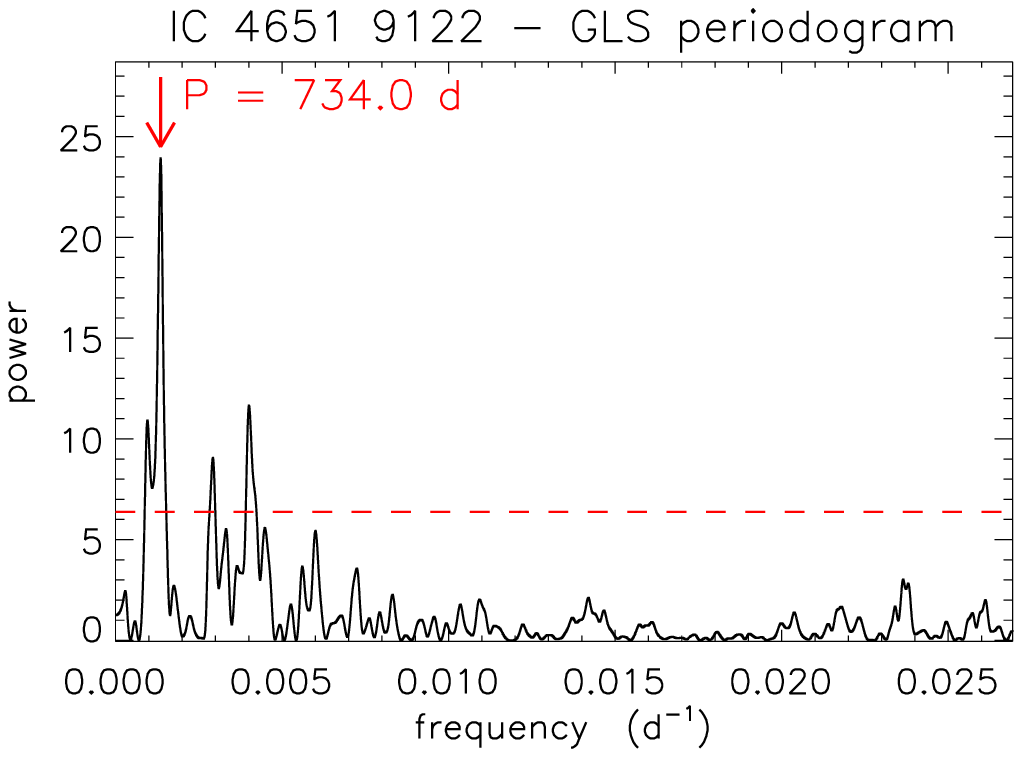}
   \includegraphics[width=\columnwidth]{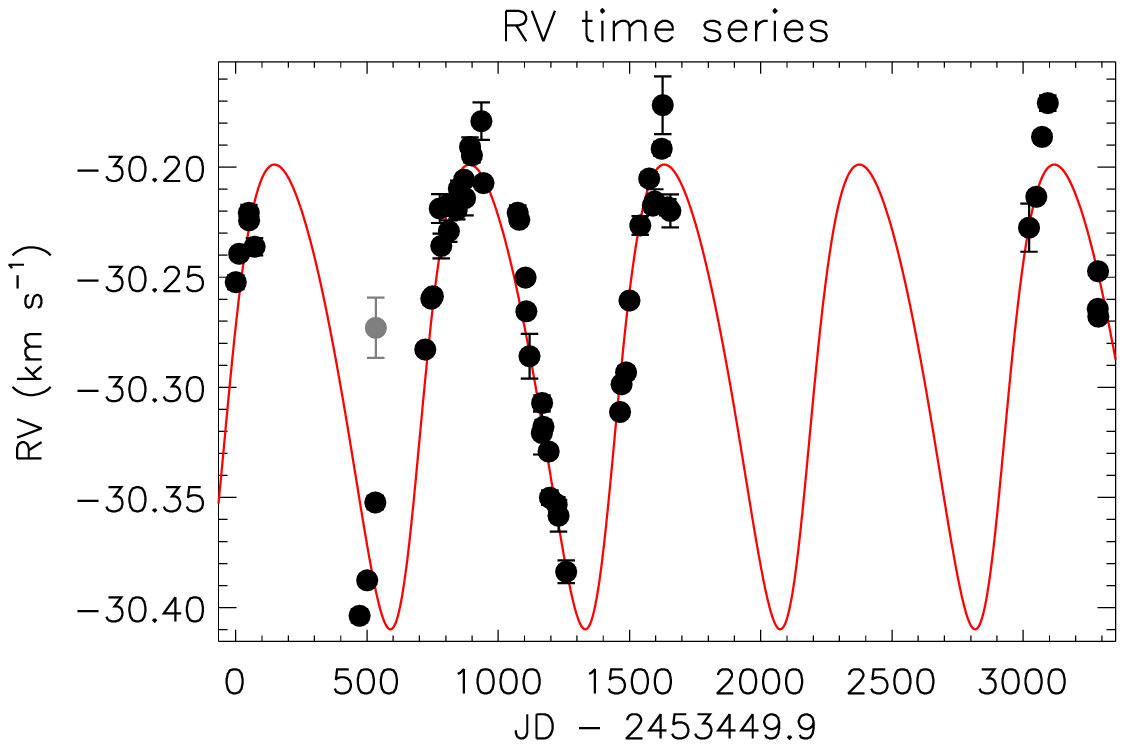}
   \includegraphics[width=\columnwidth]{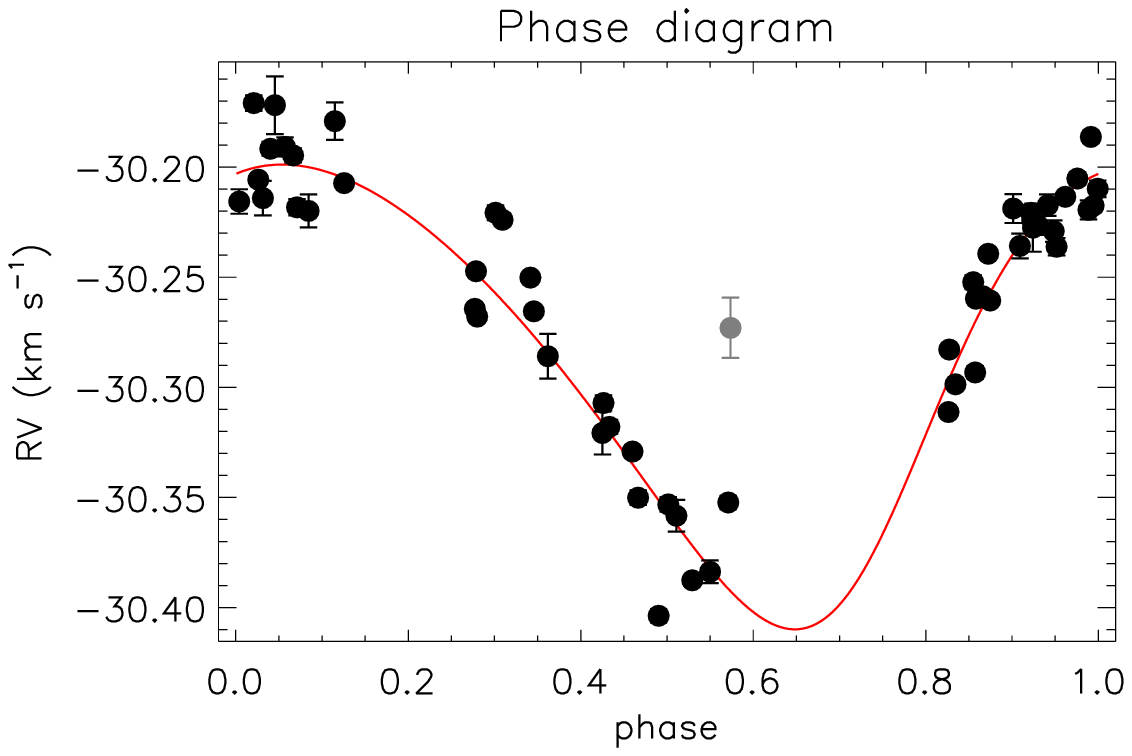}
   \caption{RV analysis of IC~4651~9122. \textit{Top panel:} GLS periodogram of the RV time series showing the most prominent peak period. The red horizontal dashed line illustrates the 1\% FAP level. \textit{Middle panel:} RV time series, where the black circles and error bars are the HARPS data and the red curve is the best Keplerian fit to the data. The gray colored datum is a particular outlier with a $S/N = 10.3$, namely very close to the threshold ($S/N = 10.0$) defined in this work. We discarded it in this analysis. \textit{Bottom panel:} phase diagram of the RV time series for the orbital period of the best fit (see Table~\ref{taborbprms}). The symbols are the same as in the middle panel.}
   \label{theplanet}
\end{figure}
%%%%%%%

%%%%%%%
\begin{figure}
  \centering 
   \includegraphics[width=\columnwidth]{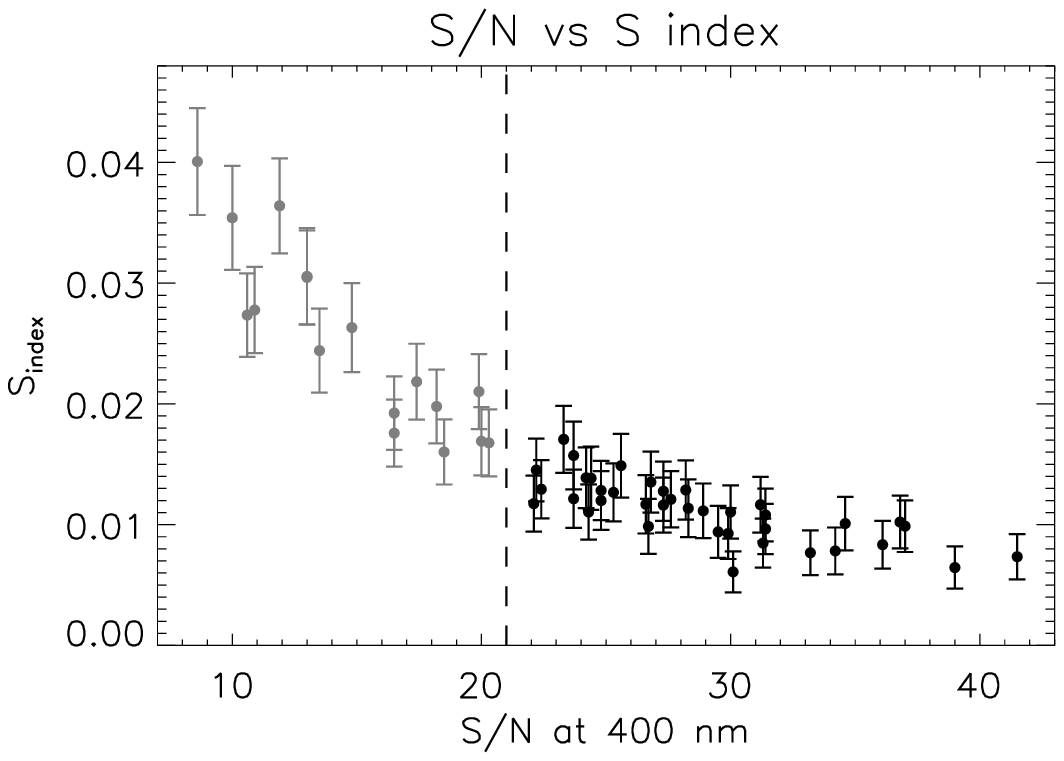}
   \caption{S~index vs. $S/N$ at 400 nm for the HARPS observations of IC~4651~9122. For $S/N < 21$, the S~index shows a strong trend with $S/N$ that is not reliable.}
   \label{snrvssindex}
\end{figure}
%%%%%%%

Even if we avoided spectroscopic binaries in our sample (based on data from the literature),
binarity can still be present. We used our HARPS observations as well as a comparison with the CORAVEL data to identify potential new binaries.

The maximum $\Delta RV/2$ value induced by a planet can be estimated by considering
a 15~M$_{\rm J}$ companion in a circular 30-day period orbit (i.e., located approximately at the Roche lobe limit) around a star of 2.0~M$_{\odot}$ (minimum stellar mass of our sample).
Such a system produces an orbital stellar semi-amplitude $K$ of $\sim$0.6~km/s, and so any targets with semi-amplitude higher than this should be associated with binary candidates.
There is one star that shows a high $\Delta RV/2$ of 764 m/s in our HARPS data. We then included this star, NGC 5822 201, in the list of binary candidates, flagged in Table~\ref{tablemain} as ``[B]'', and it is analyzed in detail in Sect.~\ref{sectheplanet}.
Apart from NGC 5822 201,
the highest $\Delta RV/2$ in our HARPS data is  198~m/s, so we have no indications of other binaries within the HARPS time series.
However, long-period binaries can be identified from a comparison between HARPS and CORAVEL.
The minimum gap between the CORAVEL (from year 1978 to 1997) and HARPS (from 2005 to 2015) observations is seven years.

Figure~\ref{histrvdiff} shows the distribution of the difference between the HARPS and CORAVEL RV data, $\left<{RV}\right> - RV_{\rm M08}$, computed by taking the average RV from each instrument.
This distribution can be well-fitted with a gaussian, from which we derive the global offset between these two instruments (the gaussian center of~131~m/s, with a $\sigma$ of~129~m/s).
The binary stars likely lie out from the peak of this distribution.
However, RV in each cluster may depend on several factors, which include stellar effective temperature, gravity, metallicity, and other systematic effects. We therefore opted for a more refined selection criteria by computing the offset between the two instruments for each cluster, instead of using the overall distribution of Fig.~\ref{histrvdiff}, for the selection of the binary candidates.
This offset was estimated by taking the median of $\left<{RV}\right> - RV_{\rm M08}$ when having three or more stars, or the smallest variation of $\left<{RV}\right> - RV_{\rm M08}$ when having only two stars.
Stars that deviate from the cluster offset by more than 0.7~km~s$^{-1}$ are considered as binary candidates. The value of 0.7~km~s$^{-1}$ represents the 5$\sigma$ distribution of Fig.~\ref{histrvdiff}. A conservative choice is justified by the consideration that such a deviation shall account also for stellar intrinsic RV variability and uncertainties in the CORAVEL measurements (typically 0.3~km~s$^{-1}$ per CORAVEL observation). In addition, 0.7~km~s$^{-1}$ is larger than the signal expected by a planet, as computed above. From Fig.~\ref{histrvdiff}, it is clear that at least eight stars have RV differences above 1.0~km~s$^{-1}$.

Using the above methodology,  13 more binary candidates were identified, flagged as ``B'' in Table~\ref{tablemain}.
These binaries were observed with HARPS only for the years 2013--2015,
explaining the reason why their variability was not detected from the HARPS data alone.
The total of  14 binaries (``B'' and ``[B]'' flags) were then removed from our sample to obtain a subsample of  101 likely single stars.
This subsample is considered in the following section, which is 
dedicated to the identification of planet-host candidates.

Table~\ref{tabif} summarizes the selection of the single-star candidates, based on the CORAVEL versus HARPS analysis described above. The number of targets for each cluster ($N_i$), the corresponding number of single star candidates ($N_f$), the instrumental RV offset ({\it Offset}), and the RV standard deviation ($\sigma$) are provided. The {\it Offset} and $\sigma$ values were computed in two steps, before and after the removal of the binaries, and the table shows the final values.
The typical offset for each cluster lies around 100--300 m/s, which is compatible with the global offset illustrated in Fig~\ref{histrvdiff}. An atypical case is NGC~2345 with an offset of $-529$~m/s, which deviates strongly from the other clusters. We cannot confidently explain the reason for this deviation.

%%%%%%%%%%%%%%%%%%%%%%%%%%%%%%%%%%%%%%%%%%%%%%%%%%%%%%%%%
\subsection{Planet-host candidates}\label{sectcandidates}

After the removal of the binaries, our subsample of likely single stars comprises  769 observations of  101 targets.
Most targets have a small number of observations and cannot be used for a proper time-series analysis and orbit determination.
We therefore selected the best planet-host candidates based on the work of \citet{hek08}, as explained in Sect.~\ref{selmethod}, with the basic criterion defined by Eqs. (25)--(27) of \citet{tro16}.

Eq.~(25) of \citet{tro16} can be represented in log-log scale by:
\begin{equation}
  \hspace{0.5cm}\log(\Delta RV_{\rm trend}/2 {\rm ~[km/s]~}) = a + b\log g,
\end{equation}
where the parameters $a = \log 2$ and $b = \frac{1}{3}\log 0.015$ reproduce the trend of \citet{hek08}.
This trend was estimated using targets with typically 20--100 observations, which \text{provide} more reliable $\Delta RV/2$ measurements than for our sample. Most of our objects have a low number of observations and this may bias the level of $\Delta RV_{\rm trend}/2$. Hence, we computed a fit to our sample by fixing the slope $b$ at the value found in \citet{tro16} and leaving $a$ as a free parameter.
We only considered for the fit the 64 objects with spectroscopic $\log g$ (see Sect~\ref{selmethod}). The best fit was obtained with $a = -0.172$~dex.

Figure~\ref{loggvsrms} depicts $\Delta RV/2$ versus $\log g$ for our subsample of  101 single star candidates.
The intrinsic variability (stellar jitter) trend is illustrated by the black solid line, the $3\times$ level is given by the black dotted line, and the 3$\times$ typical RV error is depicted by the grey dashed line.
There are  11 stars lying above both the $3\times$ trend level and the $3\times$ typical RV error, which are the best planet-host candidates.
The candidates with at least nine observations -- namely those chosen to be analyzed in more detail from their time series (see Sect.~\ref{secnobs}) -- are labeled A to D.
Some stars may rise above the threshold and become planet-host candidates when having more observations.

Table~\ref{tabcounts} provides an overview of the planet and binary candidates for all the open clusters in our program sorted by their turnoff masses $M_{\rm TO}$. These were estimated from CMD\footnote{\url{http://stev.oapd.inaf.it/cgi-bin/cmd}} Web Interface isochrones corresponding to the age and metallicity of each cluster. The parameters $\log(t)$, $[Fe/H]$, $e(B-V)$, and $\mu$ stand for their age, metallicity, reddening, and distance modulus respectively. The table also contains $N_{\rm *}$, $N_{\rm p}$, and $N_{\rm b}$, which are the number of stars belonging to each cluster of our final sample described in Sect.~\ref{finalsample}, the number of planet-host candidates identified in this work, and the number of binary candidates, respectively. Four out of the 29 open clusters observed have no targets yet with at least two effective observations.

The number of observations in our sample is still very limited to provide a proper census of planet hosts in young open clusters. For now, two clusters, IC~4756 and IC~2714, have three planet-host candidates each. This could indicate a high planet occurrence rate for these clusters or the planet-host candidates may be false positives.
These two clusters lie in a relatively narrow $M_{\rm TO}$ range around $\sim$2.5--3.1~M$_{\odot}$, whereas the clusters outside this range have either zero or one candidate.
The larger number of candidates within this range qualitatively agrees with recent studies which show that the planet occurrence rate as a function of stellar mass has a maximum for giant host stars around $\sim$2~M$_{\odot}$ \citep[e.g.,][]{ref15}.
These studies also claim that the occurrence rate drops rapidly for higher masses.
Indeed, we have only one candidate among the sample of the 40 most massive stars
($M_{\rm TO} > 3.1$~M$_{\odot}$).
Such a low incidence of planets in massive stars can be understood within a scenario in which strong winds from high-mass stars may generate competing timescales between disk dissipation and planet formation \citep[e.g.,][]{ken09,rib15}.
However, we shall note that some observational biases may reduce the planet detectability when increasing the host mass \citep[e.g.,][]{jon14}.
For instance, some of our planet-host candidates would produce orbital semi-amplitudes below the 3$\times$-trend line of Fig.~\ref{loggvsrms} if they were observed around more massive stars (see Sect.~\ref{secmp} for a specific example). In addition, more luminous giants tend to have a larger intrinsic stellar noise, and this also introduces bias.

%%%%%%%
\begin{figure*}
  \centering 
   \includegraphics[width=\columnwidth]{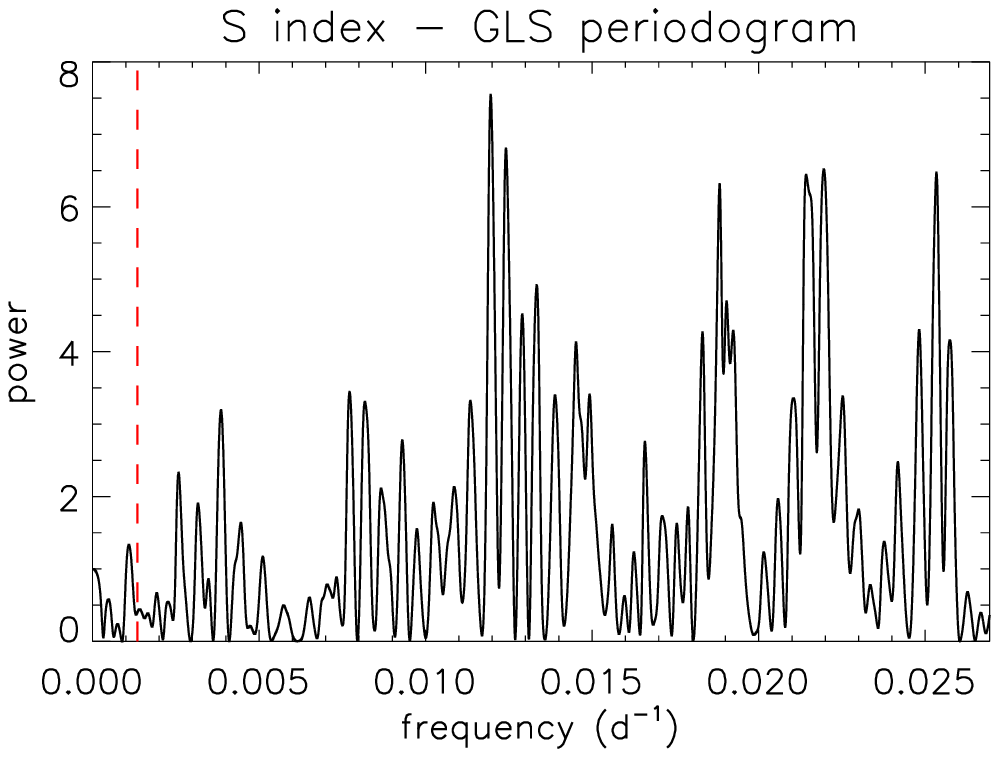}
	 \includegraphics[width=\columnwidth]{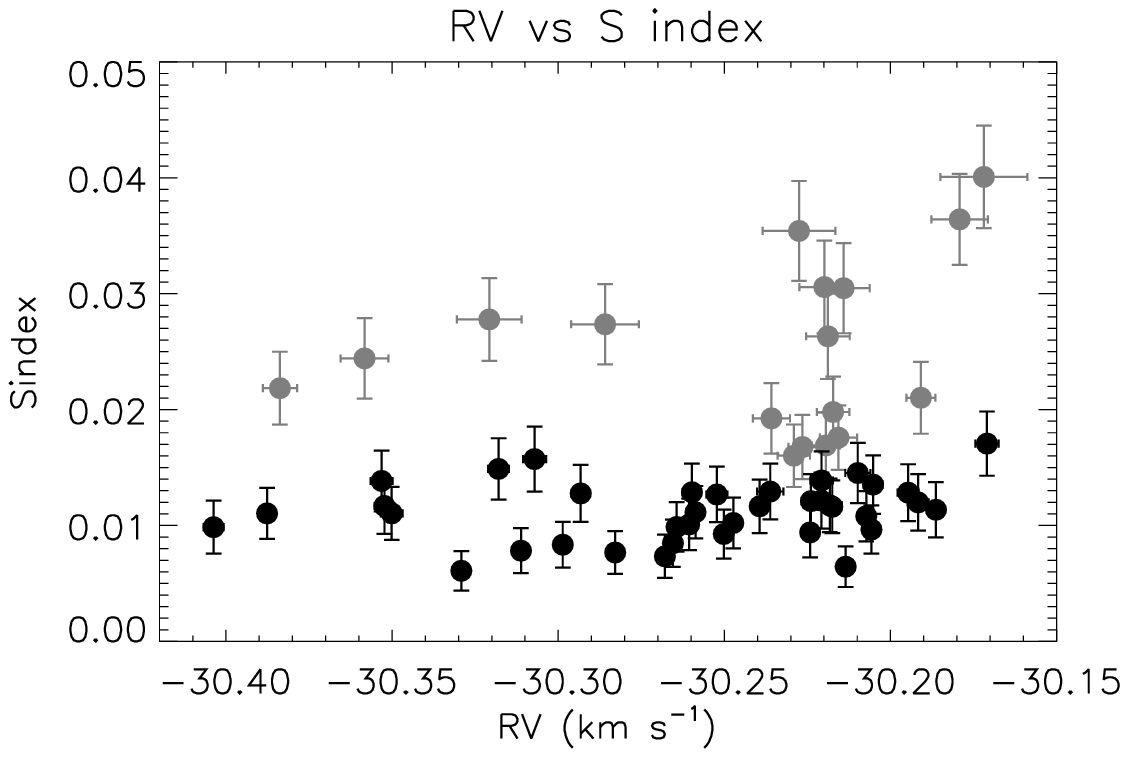}
   \includegraphics[width=\columnwidth]{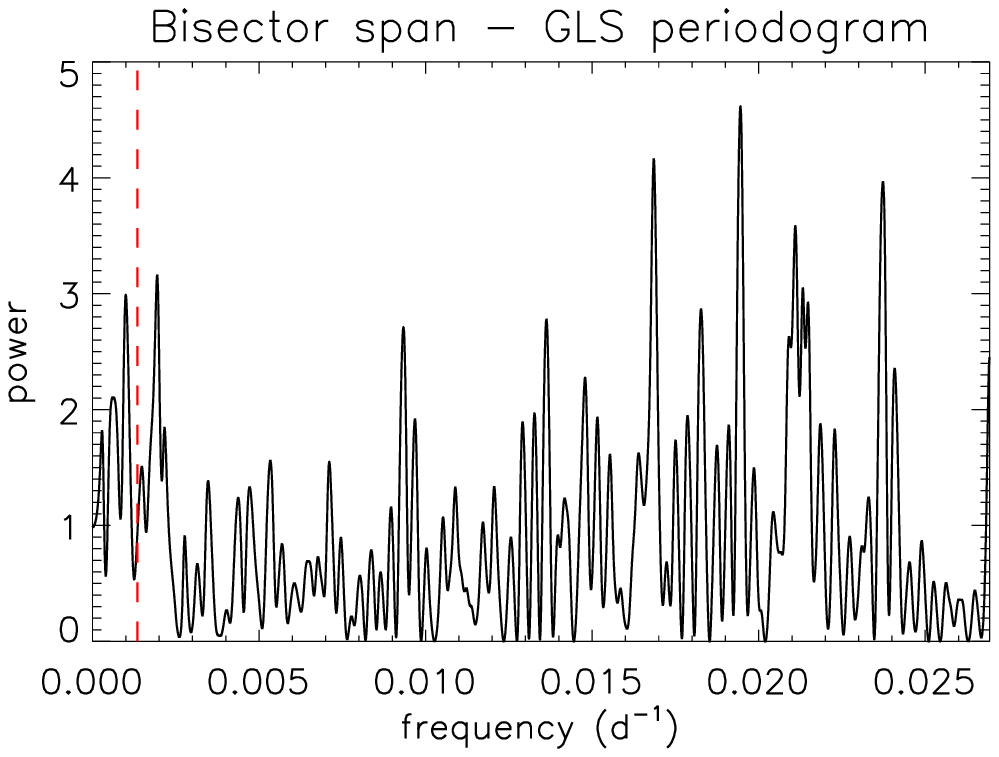}
   \includegraphics[width=\columnwidth]{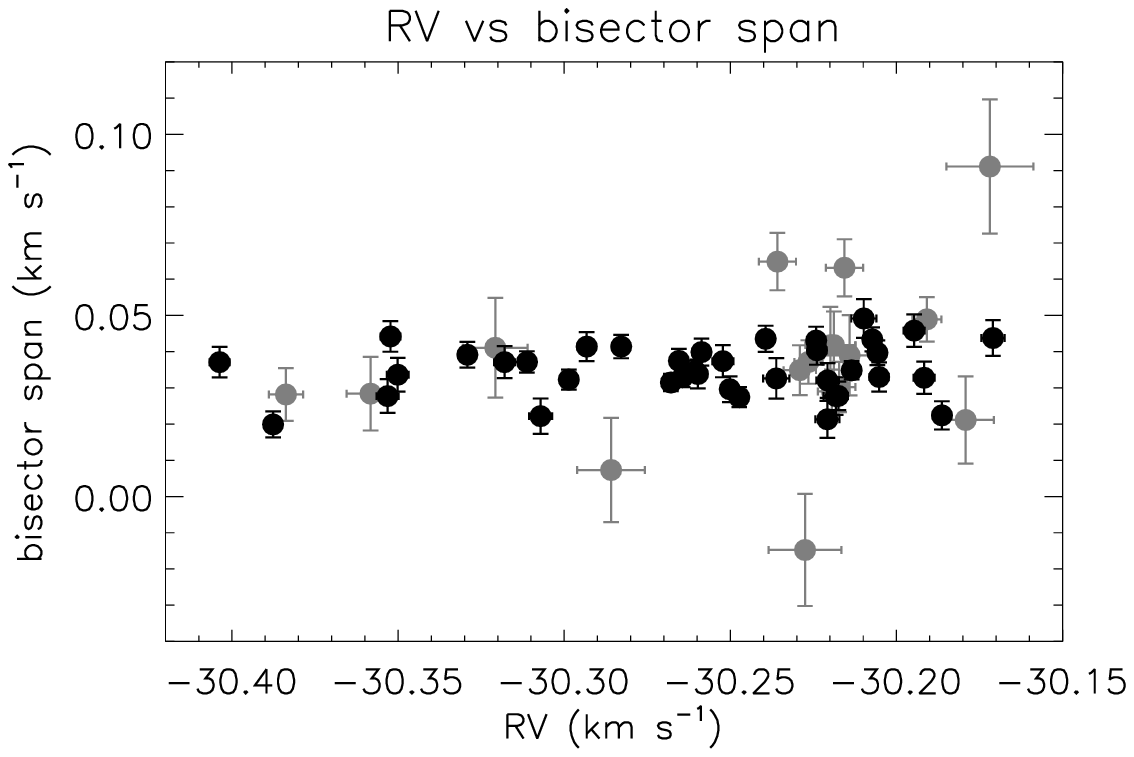}
   \caption{Analysis of activity proxies for IC~4651~9122. \textit{Left panels:} GLS periodograms of the S~index and of the bisector span time series by considering the data subset with $S/N \geq 21$. The red vertical dashed line illustrates the orbital period of the Keplerian fit described in Table~\ref{taborbprms}. \textit{Right panels:} correlation between RV and each activity proxy, where the data is split into $S/N < 21$ (gray circles) and $S/N \geq 21$ (black circles). The Pearson correlation coefficient is  0.20 for RV versus S~index and  0.15 for RV versus bisector span when considering the data subset with $S/N \geq 21$.}
   \label{testactiv}
\end{figure*}
%%%%%%%

%%%%%%%
\begin{figure*}
  \centering 
   \includegraphics[width=\columnwidth]{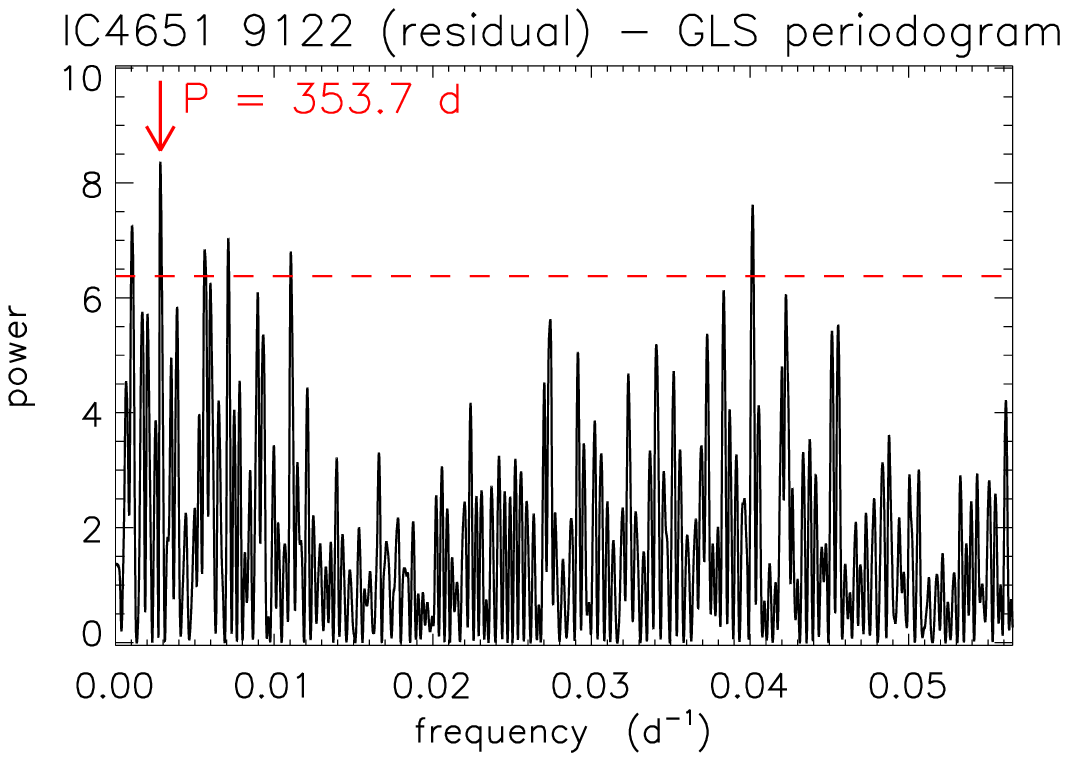}
	 \includegraphics[width=\columnwidth]{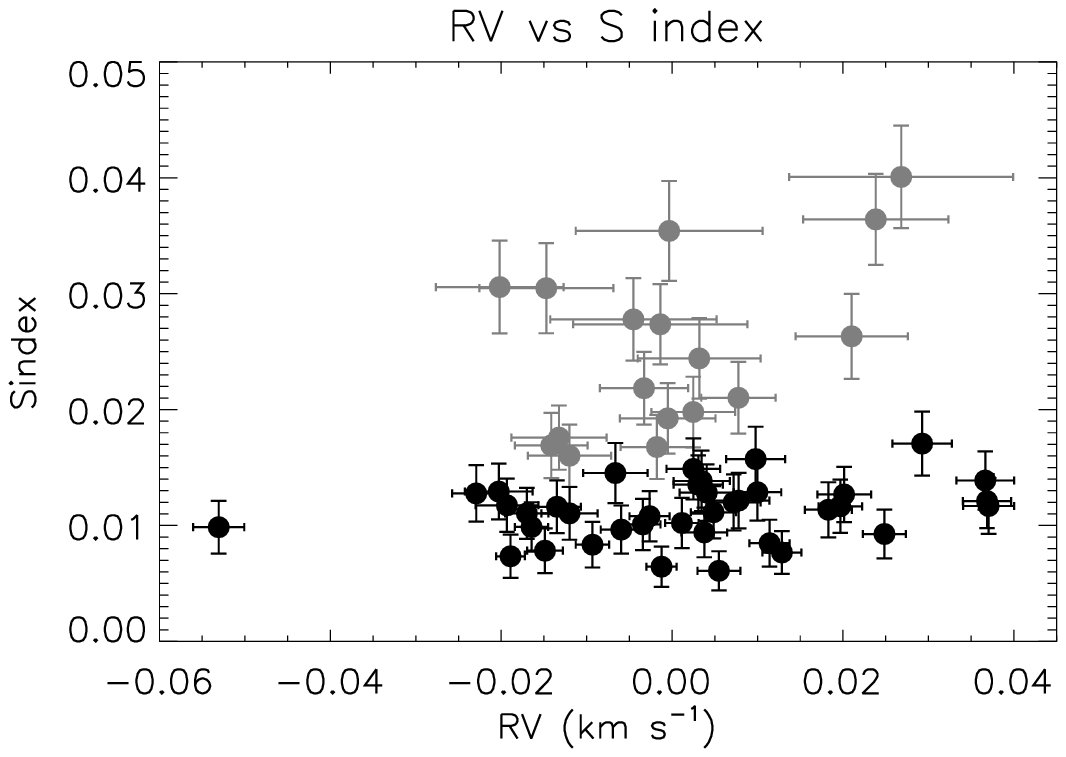}
   \includegraphics[width=\columnwidth]{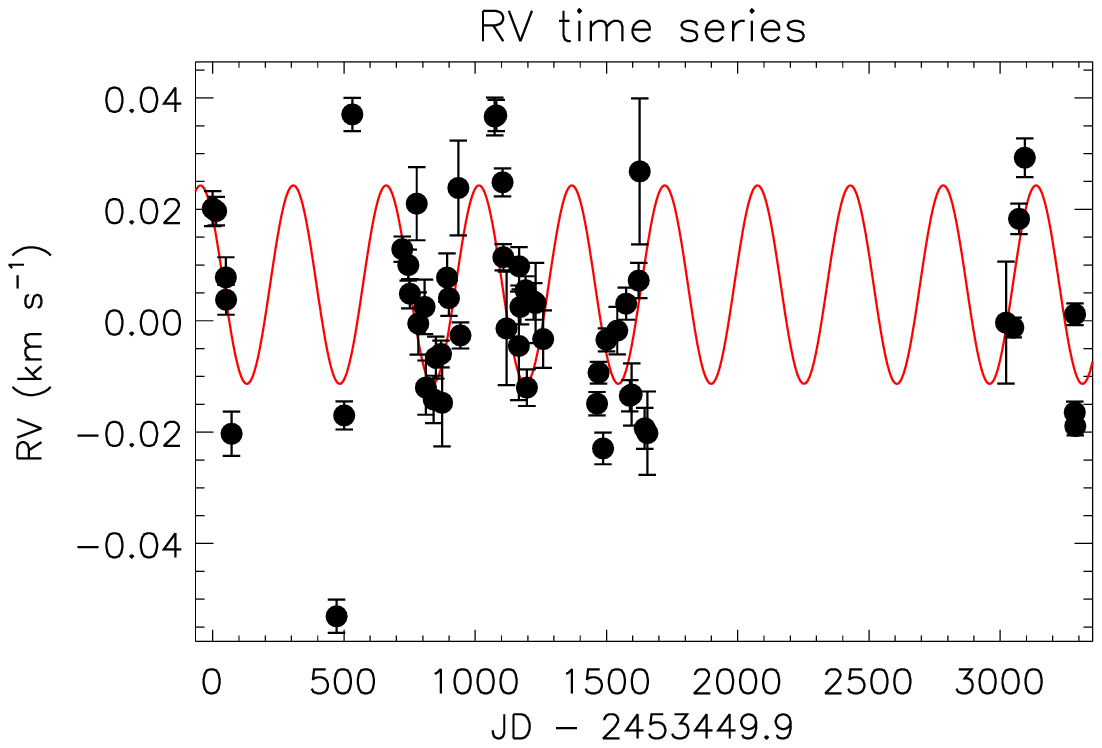}
   \includegraphics[width=\columnwidth]{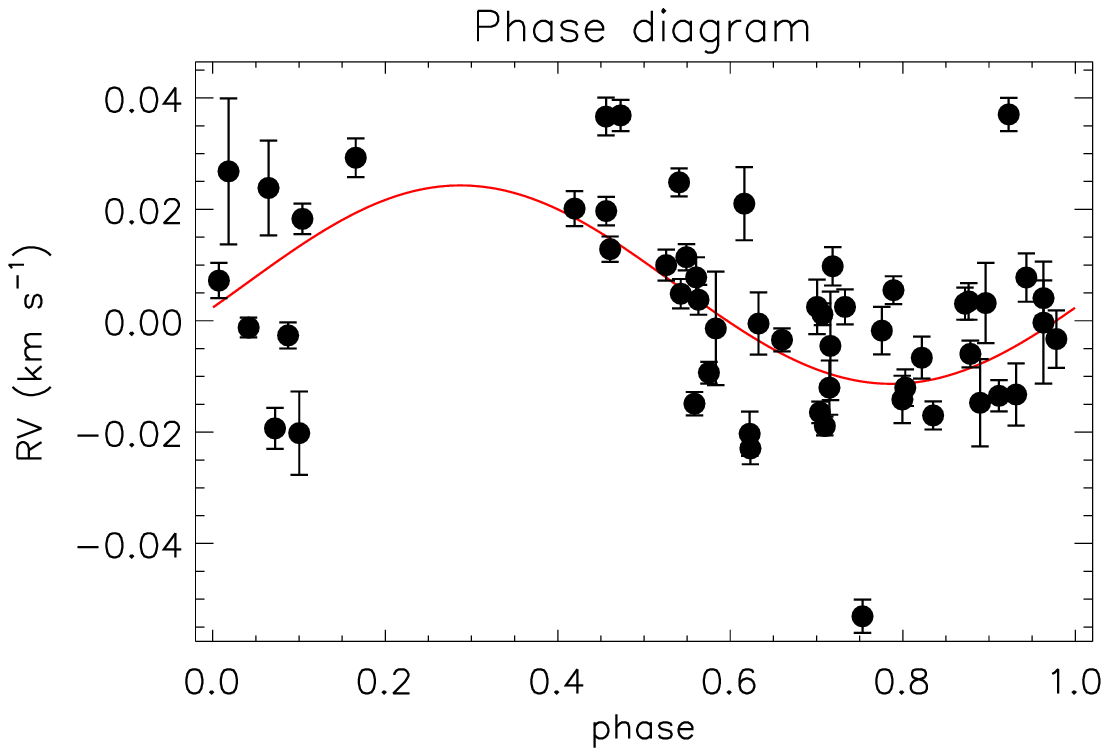}
   \caption{RV and activity proxy analyses for the residual of the RV time series of IC~4651~9122 after removing the best Keplerian fit. The GLS periodogram (\textit{top left panel}), the RV versus S~index correlation (\textit{top right panel}), the RV time series (\textit{bottom left panel}) and the RV phase diagram (\textit{bottom right panel}) follow similar definitions to those described in Figs.~\ref{theplanet} and \ref{testactiv}. The Pearson correlation coefficient is  0.27 for RV versus S~index and  0.15 for RV versus bisector span when considering the data subset with $S/N \geq 21$.}
   \label{residual}
\end{figure*}
%%%%%%%

%%%%%%%%%%%%%%%%%%%%%%%%%%%%%%%%%%%%%%%%%%%%%%%%%%%%
\subsection{Long-term variations in IC~2714 stars}
\label{sectimeseries}

Targets labeled~A, B, and~C in Fig.~\ref{loggvsrms} have at least nine observations and lie above the $3\times$ level of the $\log g$ versus RV trend. Hence, these are good planet-host candidates and, theoretically, they have enough data for a time series analysis. We analyzed Generalized Lomb-Scargle (GLS) periodograms \citep{zec09} for these RV time series to look for orbit-related signals. The periodograms provide periods with false alarm probability (FAP) less than 1\%, but Keplerian fits to the data provide doubtful or ambiguous solutions.

These targets all belong to the same cluster, IC~2714, and their RV time series show linear trends of a few (5--10) m/s/yr, as shown in Fig~\ref{timeseries}.
No pulsation or rotational modulation with such a long period is expected in the evolutionary stage of these stars (somewhat between the RGB-base and the red clump; e.g., \citealt{del16}).
Our S~index measurements also do not show any conclusive correlation, so longer time span observations are needed to verify the origin of the RV variation, including the possibility of a substellar companion.

%%%%%%%%%%%%%%%%%%%%%%%%%%%%%%%%%%%%%%%%%%%%%%%%%%%%%%%%
%\subsection{IC~4651~9122b: a new Jupiter orbiting a giant star}
\subsection{Orbital solutions for IC~4651~9122b and NGC~5822~201B}
\label{sectheplanet}

The star labeled~D, IC~4651~9122, has a very clear RV periodic variation.
Figure~\ref{theplanet} shows a set of standard analyses performed for this star.
The GLS periodogram (top panel) shows a strong peak with a FAP~$< 10^{-9}$ corresponding to a peak of about 2~yr.
A Keplerian model fits well with the observed data, as shown in the middle and bottom panels,
suggesting the presence of a planet, namely IC~4651~9122b.

From Fig.~\ref{loggvsrms}, the star is expected to have an intrinsic variability (jitter) of $\sim$20~m~s$^-1$.
To be conservative, an upper limit for this jitter is of $\sim$60~m~s$^-1$, namely the 3$\times$ trend.
We used the RVLIN and BOOTTRAN codes\footnote{\url{http://exoplanets.org/code/}} \citep{wri09,wan12} 
to calculate the orbital parameters by testing different jitter levels from~10 to~60~m~s$^-1$.
In a first test, we performed a Monte Carlo approach by computing independent Keplerian fits with random fluctuations to the RV data within a certain jitter level added in quadrature to the RV errors.
For verification, we used the bootstrapping method from the BOOTTRAN code.
The solutions were rather stable for or all the tests, including for an upper jitter level of 60~m~s$^-1$.
The orbital parameters obtained from our Monte Carlo approach for the most likely stellar jitter level of 20~m/s are given in Table~\ref{taborbprms}.

Although the RV periodic variation in Fig.~\ref{theplanet} is obvious, we use the S~index and the bisector span to establish the nature of the RV periodicity.
We have to consider a data subset with high $S/N$ for the S~index because this parameter is not reliable at low $S/N$. Figure~\ref{snrvssindex} shows a systematic increase of the S~index with decreasing $S/N$, which occurs because the CaII H \& K lines become dominated by noise.
The bisector also loses confidence at low $S/N$.
Hence, we split the data into lower and higher $S/N$ regimes, namely $S/N < 21$ and $S/N \geq 21$ (at 400~nm), for a proper interpretation of our results.
These regimes are illustrated in our analysis by the gray and black circles, respectively.

Figure~\ref{testactiv} displays the activity proxy tests. The GLS periodograms (left panels) show no confident period related with the orbit of the planet, and the RV versus activity proxy diagrams (right panels) have low correlations.
A small subset of data where the S~index seems to increase with increasing RV (gray circles in the top-right panel) is
not to be trusted, as the $S/N$ of the data is too low.
Overall, this analysis shows no association between the activity proxies and RV, thus supporting the orbital origin for the main RV variation.

%%%%%%%
\begin{figure}
  \centering 
	 \includegraphics[width=\columnwidth]{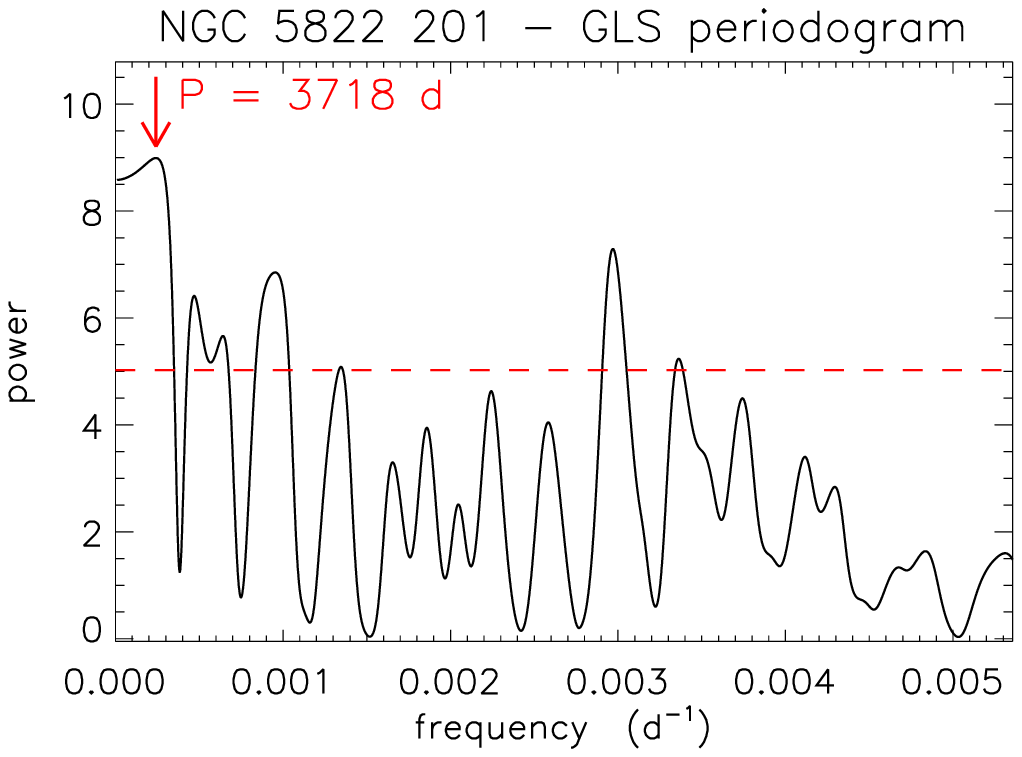}
   \includegraphics[width=\columnwidth]{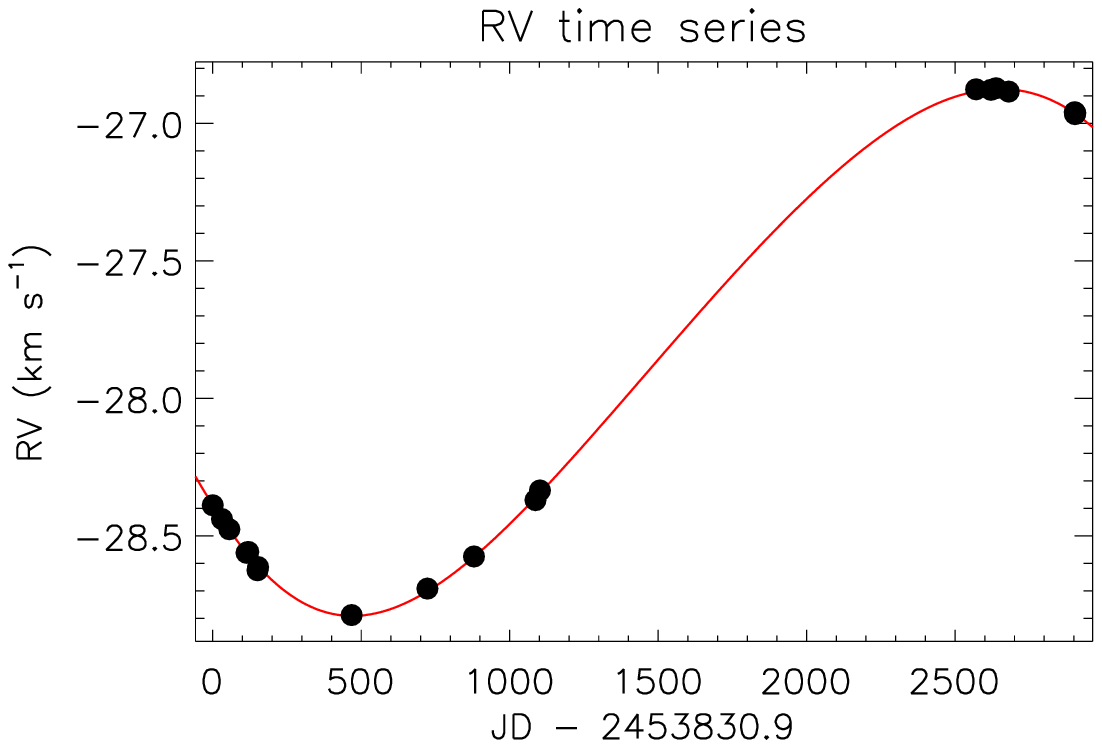}
   \caption{RV time series analysis for NGC~5822~201. \textit{Top panel:} GLS periodogram. \textit{Bottom panel:} RV time series. Symbols follow the same definitions as in Fig.~\ref{theplanet}.}
   \label{solbinary}
\end{figure}
%%%%%%%

%%%%%%%
\begin{table}
   \caption{Orbital parameters for the new giant planet IC~4651~9122b.}
   {\centering
   \begin{tabular}{l c}
     \hline\hline
       Orbital period (d) & $734.0 \pm 8.1$ \\
       Minimum mass (M$_{\rm J}$) & $6.3 \pm 0.5$ \\
       Semi-major axis (AU) & $2.038 \pm 0.039$ \\
       Eccentricity & $0.18 \pm 0.09$ \\
       RV semi-amplitude (m s$^{-1}$) & $89.5 \pm 6.8$ \\
       Argument of periastron (deg) & $118.5 \pm 60.7$ \\
       Time of periastron (JD) & $2454605.6 \pm 175.0$ \\
     \hline
     \end{tabular} \\ }
  \label{taborbprms}
\end{table}
%%%%%%%

%%%%%%%
\begin{table}
   \caption{Orbital parameters for the binary companion NGC~5822~201B.}
   {\centering
   \begin{tabular}{l c}
     \hline\hline
       Orbital period (d) & $3718 \pm 325$ \\
       Minimum mass (M$_{\odot}$) & $0.112 \pm 0.005$ \\
       Semi-major axis (AU) & $6.497 \pm 0.098$ \\
       Eccentricity & $0.15 \pm 0.07$ \\
       RV semi-amplitude (m s$^{-1}$) & $960.1 \pm 18.5$ \\
       Argument of periastron (deg) & $100.5 \pm 39.5$ \\
       Time of periastron (JD) & $2453670.9 \pm 936.0$ \\
     \hline
     \end{tabular} \\ }
  \label{tabbinprms}
\end{table}
%%%%%%%

We also analyze the residual of the observed RV data after removing the best Keplerian fit for IC~4651~9122, as shown in Fig~\ref{residual}. There is a signal which is still noticeable after subtraction whose nature is yet to be determined.
The GLS periodogram provides a prominent peak period of about 1~yr, which is half of the 2-yr period of IC~4651~9122b.
However, systematic effects, such as 1-yr seasonal variation, may contribute to this period, which does not survive when removing some data points at random.
The $\Delta RV/2$ level of this residual (see Fig.~\ref{loggvsrms}) lies below the region we defined for planet-host candidates,
so this signal may not be caused by a second planet. The S~index in Fig.~\ref{residual} (top right panel) does not exclude the possibility of a second planet, since it shows no significant correlation with RV.

Finally, we analyzed the time series of NGC~5822~201, identified in Sect.~\ref{binaries} as a binary candidate.
Fig.~\ref{solbinary} shows the periodogram of the RV time series in the top panel and the RV time series in the bottom panel.
The orbital nature of this signal is very clear
and the best Keplerian fit parameters computed from our Monte Carlo approach are given in Table~\ref{tabbinprms}.
For proper error calculations, we assumed a jitter of~45~m~s$^{-1}$ for the primary star based on its $\log g$ value of~$2.85 \pm 0.08$~dex and on the trend curve of Fig.~\ref{loggvsrms}.

%%%%%%%%%%%%%%%%%%%%%%%%%%%%%%%%%%%%%%%%%%%%%%%%%%%%%%%%
\subsection{Possible orbital parameters for the planet candidates}
\label{secmp}

We provide in this section an overall discussion of the planet-host candidates by considering possible orbital solutions in case the planets were confirmed.
The minimum planet mass, $m\sin i$, is computed from the RV equation, where the stellar mass and the RV~semi-amplitude are assumed to be $M_* \simeq M_{\rm TO}$ and $K \simeq \Delta RV/2$, respectively.
The orbital period $P$ and eccentricity $e$ are unknown for any planet candidate (except for the confirmed planet~IC~4651~9122b).
We thus assume a low eccentricity ($\lesssim$0.3) for possible planets, whereas, as far as orbital periods are concerned, a more detailed discussion is required. That discussion is presented below, being mostly based on a visual inspection of the~RV~time series
with a limited number of observations.

Rough orbital parameters can be suggested for the IC~2714 planet-host candidates based on Fig.~\ref{timeseries} from Sect~\ref{sectimeseries}. In general, the long-term RV variations of these candidates would fulfil at most half a cycle of hypothetical orbits for all cases. Orbital periods should therefore be as long as at least twice the total time span (of $\sim$3200--3700~d) of the HARPS observations. From this assumption, hypothetical planets would lie more than $\sim$10~AU away from their host stars and would have masses greater than $\sim$10~M$_{\rm J}$.
These candidates are good examples for illustrating the detectability bias mentioned in Sect.~\ref{sectcandidates}: they would produce RV semi-amplitudes about 30\% smaller, thus below the 3$\times$-trend line of Fig.~\ref{loggvsrms}, if they were observed around a 6~M$_{\odot}$ star.

Limited observations of the remaining candidates show, except for the most massive one (NGC~2925~108, with 5.9~M$_{\odot}$), noticeable RV variations likely within short time spans. Such a signature is compatible either with possible presence of close-in planets with a few Jupiter masses or with intrinsic stellar signal. The most massive candidate has six observations spread over a $\sim$700~d time span showing likely a long-term variation similar to the IC~2714 candidates, thus indicating presence of a possible long-period and massive planet.

Overall, an interesting aspect is the close similarity between the RV variations of the planet-host candidates in IC~2714. If these variations were due to planetary companions, their preliminary orbital parameters would indicate presence of rather massive planets around massive stars. Such a trend qualitatively agrees with a commonly proposed scenario where more massive stars would be formed with correspondingly more massive protoplanetary disks that would yield more massive planets \citep[e.g.,][]{ida05,ken08}.
Finally, the NGC~2925~108 candidate is another interesting case because, if confirmed, it may extend planet detection in open clusters over a broader stellar mass range where a low planet incidence has been observed (see Sect.~\ref{sectcandidates}).

%%%%%%%%%%%%%%%%%%%%%%%%%%%%%%%%%%%%%%%%%%%%%%%%%%%%%%%%%%%%%%%%%%%%%%%%%%%%
\section{Conclusions} \label{conclusions}
%%%%%%%%%%%%%%%%%%%%%%%%%%%%%%%%%%%%%%%%%%%%%%%%%%%%%%%%%%%%%%%%%%%%%%%%%%%%

We present the first results of a long-term survey where we look for massive exoplanets orbiting intermediate-mass ($\sim$2--6~M$_{\odot}$) giant stars belonging to 29 open clusters.
This survey aims to provide in the future, following the collection of more data, a census of a diversity of stellar and planetary environments with detailed physical descriptions. These will include stellar absolute physical parameters and planetary orbital solutions, based on a homogeneous set of HARPS observations.

We have identified 14 new binary candidates, by combining our observations with CORAVEL data, spanning a $\sim$30~yr (from 1978 to 2015) baseline,
despite sample pre-selection aimed to avoid binaries.
We then considered  101 single stars, among which we detected  11 planet-host candidates.
It is worth noting that  10 of the  11 candidates have masses $\lesssim 3.2$~M$_{\odot}$, and only one candidate has mass greater than this.
This agrees qualitatively with recent studies concerning the occurrence rate of massive planets as a function of the host star masses, which shows, for the case of giant stars, a peak around $\sim$2M$_{\odot}$ and a low rate for higher masses \citep[e.g.,][]{ref15}.
We however warn that our selection method has an intrinsic bias against more massive stars.

Three of the planet-host candidates belong to the same cluster, IC~2714, and show common behavior, which is long-term RV variation that cannot be resolved with the current observations.
More observations are needed to verify whether they are induced by substellar companions.

One planet-host candidate, IC~4651~9122, shows very clear RV periodic variation. Time series analysis and tests of activity proxies confirmed this star has a giant planet companion, namely IC~4651~9122b, with a minimum mass of 6.3~M$_{\rm J}$ and a semi-major axis of 2.0~AU.
There is a residual signal that may have a physical origin, but it also requires more observations for proper interpretation.
Finally, one of the binary candidates, NGC~5822~201, also has enough data to study in further detail.
The companion, NGC~5822~201B, has a very low minimum mass of 0.11~M$_{\odot}$ and a semi-major axis of 6.5~AU, which is comparable to the Jupiter distance to the Sun.

The number of known sub-stellar objects around giants is still rather small. Brown dwarfs orbiting Sun-like stars seem to become less frequent when the mass of the star increases, whereas planets become more frequent \citep[e.g.,][]{gre06}. This behavior may be different for more massive or giant stars, as suggested in some studies \citep[e.g.,][]{lov07} and indicated qualitatively in Sect.~\ref{sectcandidates}.
Overall, our small sample of planet-host candidates can extend this study to more massive stars in relation to previous works. It indicates a possible dependence of the planet incidence upon the stellar mass, as well as some relation between host star mass and planetary mass, that are qualitatively compatible with theoretical and observational studies.
Observing more giant stars in clusters may therefore provide essential information to better understand those distributions among other aspects.

%%%%%%%%%%%%%%%%%%%%%%%%%%%%%%%%%%%%%%%%%%%%%%%%%%%%%%%%%%%%%%%%%%%%%%%%%%%%
\begin{acknowledgements}
%%%%%%%%%%%%%%%%%%%%%%%%%%%%%%%%%%%%%%%%%%%%%%%%%%%%%%%%%%%%%%%%%%%%%%%%%%%%

Research activities of the Observational Astronomy Board of the Federal University of Rio Grande do Norte (UFRN) are supported by continuous grants from the CNPq and FAPERN Brazilian agencies.
ICL acknowledges a Post-Doctoral fellowship at the European Southern Observatory (ESO) supported by the CNPq Brazilian agency (Science Without Borders program, Grant No. 207393/2014-1).
BLCM acknowledges a PDE fellowship from CAPES.
SA acknowledges a Post-Doctoral fellowship from the CAPES Brazilian agency (PNPD/2011: Concess\~ao Institucional), hosted at UFRN from March 2012 to June 2014.
GPO acknowledges a graduate fellowship from CAPES.
Financial support for CC is provided by Proyecto FONDECYT Iniciaci\'on a la Investigacion 11150768 and the Chilean Ministry for the Economy, Development, and Tourism's Programa Iniciativa Cient\'ifica Milenio through grant IC120009, awarded to Millenium Astrophysics Institute.
DBF acknowledges financial support from CNPq (Grant No. 306007/2015-0).
LP acknowledges a distinguished visitor PVE/CNPq appointment at the Physics Graduate Program of UFRN in Brazil and thanks to DFTE/UFRN for hospitality. We also acknowledge the Brazilian institute INCT INEspa\c{c}o for partial financial support.
Finally, we warmly thank the anonymous referee for very constructive comments.

\end{acknowledgements}


\end{document}